\documentclass[twocolumn,aps,prl,showpacs,amsmath,superscriptaddress,longbibliography,notitlepage]{revtex4-1}
\usepackage{amssymb}
\usepackage{mathrsfs}
\usepackage{graphicx}
\usepackage{float}
\usepackage[caption=false]{subfig}
\usepackage[pdftex,colorlinks,linkcolor={red!75!black},citecolor={red!75!black},urlcolor={blue!75!black}]{hyperref}
\usepackage{epstopdf}
\usepackage{xcolor}
\usepackage{bm}
\usepackage{soul}

\newcommand{\clr}{\color{red!75!black}}

\newcommand{\Rnum}[1]{\uppercase\expandafter{\romannumeral #1\relax}}

\usepackage{soul}
\usepackage{changes}

\usepackage{verbatim}

\begin{document}

\title{Occupation-dependent particle separation in one-dimensional non-Hermitian lattices}
\author{Yi Qin}
\affiliation{Guangdong Provincial Key Laboratory of Quantum Metrology and Sensing $\&$ School of Physics and Astronomy, Sun Yat-Sen University (Zhuhai Campus), Zhuhai 519082, China}
\author{Linhu Li}\email{lilh56@mail.sysu.edu.cn}
\affiliation{Guangdong Provincial Key Laboratory of Quantum Metrology and Sensing $\&$ School of Physics and Astronomy, Sun Yat-Sen University (Zhuhai Campus), Zhuhai 519082, China}
\begin{abstract}
	We unveil an exotic phenomenon arising from the intricate interplay between non-Hermiticity and many-body physics, namely an 	occupation-dependent particle separation for hardcore bosons in a one-dimensional lattice driven by uni-directional non-Hermitian pumping. 	Taking hardcore bosons as an example, we find that a pair of particles occupying the same unit cell exhibit an opposite non-Hermitian pumping direction to that of unpaired ones occupying different unit cells.
	By turning on an intracell interaction, many-body eigenstates split in their real energies, forming separable clusters in the complex energy plane with either left-, right-, or bipolar-types of non-Hermitian skin effect (NHSE).
	The dependency of skin accumulating directions on particle occupation is further justified with local sublattice correlation and entanglement entropy of many-body eigenstates.
	Dynamically, this occupation-dependent NHSE manifests as uni- or bi-directional pumping for many-body initial states, allowing for spatially separating paired and unpaired particles. Similar phenomena also apply to fermionic systems, unveiling the possibility of designing and exploring novel non-Hermitian phases originated from particle non-conservation in subsystems (e.g., orbitals, sublattices, or spin species) and their spatial configurations.   
\end{abstract}

\maketitle
\emph{{\clr Introduction}.---}
Many-body interaction is known to trigger many exotic physics beyond single-particle picture,
ranging from fractional quantum Hall effect and quantum spin liquids, to dynamical fermionization,
Hilbert space framentation,
and quantum time crystal
in non-equilibrium systems.
When coupled to an external environment, quantum many-body systems can be effectively described by non-Hermitian Hamiltonians~\cite{bender1998nonH,bender2007making,Konotop2016,Ganainy2018,ashida2020non,Bergholtz2021},
which exhibit fundamentally different properties from Hermitian systems, and have arose much theoretical~\cite{Rudner2009,Esaki2011,Hu2011PRB,Henning2013,Lee2016,Leykam2017,Xu2017PRL,Shen2018,Martinez2018,Yao2018,Kunst2018,McDonald2018,Lee2019,Kawabata2019,Zhang2020,Okuma2020,li2022non,Koch2022,Budich2020,Li2021,Liu2021,Mu2022,li2022non,tai2023zoology,qin2023universal,qin2023kinked,Borgnia2020,Deng2022PRB} and experimental~\cite{Poli2015,Zeuner2015,zhen2015,Weimann2017,XiaoL2017,StJean2017,Babak2017,Zhao2018,Hengyun2018,Helbig2020,Hofmann2020,xiaoLei2020,Sebastian2020,Palacios2021,ZhangXiujuan2021} attention during recent years.
In particular, 
a major focus is the non-Hermitian skin effect (NHSE), where eigenstates massively and non-reciprocally accumulate to system's boundaries~\cite{Martinez2018,Yao2018,lin2023topological}.
In single-particle level,
a variety of NHSE has been discovered with richer non-reciprocal pumping channels,
e.g., bipolar and reciprocal NHSE with bidirectional skin accumulation~\cite{Song2019_PRL}, geometry-dependent NHSE from anisotropic non-reciprocity~\cite{Kai2022NC,Wang2022_NC,Fang2022,wang2023experimental,wan2023,YiQin2023}, and hybrid skin-topological effect with non-reciprocal pumping only applied to boundary states~\cite{LeeCH2019_PRL,li2020topological,zou2021observation,li2022gain,zhu2022hybrid,Zuxuan2023,lei2023mathcal}.
In the realm of many-body physics, many efforts have been made in investigating how NHSE affects many-body phenomena~\cite{lee2021many,Orito2022, Suthar2022, wang2023nH,qin2023nonH,likai2023,Federico2023}, or vice versa~\cite{Kawabata2022, Zhang2022, ora2022, Zhang2022PRB, Wang2022, zheng2023Ex, li2023MB, ZhuBofen2022,MusenPRB2020}. 
Intriguing many-body cluster phenomena have also be found to arise from non-Hermitian interactions~\cite{Faugno2022, Shen2022}.

In this paper, we report an emergent occupation-dependency of NHSE originated from the interplay of multiple non-reciprocal pumping channels and many-body effects of hardcore bosons.
Namely, a pair of particles occupying the same unit cell experience skin accumulation opposite  to that of unpaired particles occupying different unit cells,
thus they are separated to different ends of a one-dimensional (1D) lattice, whilst all single-particle eigenstates localize at the same end of the lattice. 
Eigenstate clusters with left-, right-, or bipolar NHSE at different real eigenenergies are thus formed when turning on a {\it Hermitian} intracell interaction.
Physically, such an enigmatic phenomenon can be understood with the destructive interference of non-reciprocal pumping channels induced by a strong intracell hopping, and its deactivation for paired particles occupying the same unit cells.
This occupation-dependency of many-body NHSE is further verified by our calculation of local sublattice correlation and entanglement entropy.
Dynamically, paired and unpaired particles are observed to be pushed to opposite directions, 
constituting a particle splitter depending on their paring conditions.

\emph{{\clr NHSE in a 1D ladder lattice}.---}
\begin{figure}[htbp]
\centering
\includegraphics[width=\linewidth]{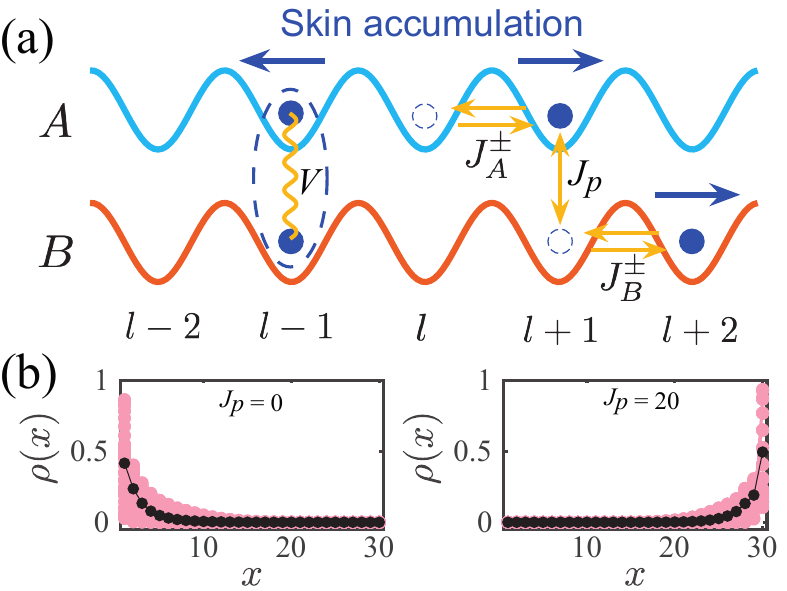}
\caption{\label{fig_Model} Schematic of a 1D ladder lattice with two sublattices in a unit cell, each represents a Hatano-Nelson model, and NHSE in its single-particle level. 
(a) The hopping amplitude to left (right) direction is $J_{A,B}^-$ ($J_{A,B}^+$). The intracell hopping strength and interaction are $J_p$ and $V$ respectively. Red arrows indicate the skin accumulation directions for 
paired and unpaired particles. (b) Distribution of single-particle eigenstates for different $J_p$.
Other parameters are $J_{A}^+=0.45, J_{A}^-=1.24$, $J_{B}^+=-0.82$,$J_{B}^{-}=-1.22, L=30$.
}
\end{figure}
As sketched in Fig. \ref{fig_Model}(a), we consider hardcore bosons loaded in a 1D ladder lattice with two sublattices $A$ and $B$ with $L$ lattice sites, described by the Hamiltonian
\begin{eqnarray}
\hat{H} &=&  \sum\limits_{l=1}^{L-1}\mathbf{\hat{\psi}}^\dagger_{l+1}
\left(\begin{array}{cc}
J^+_A & 0\\
0 & J^+_B
\end{array}\right)
\mathbf{\hat{\psi}}_{l}+\mathbf{\hat{\psi}}^\dagger_{l}
\left(\begin{array}{cc}
J^-_A & 0\\
0 & J^-_B
\end{array}\right)
\mathbf{\hat{\psi}}_{l+1}
\nonumber\\
&&+\mathbf{\hat{\psi}}^\dagger_{l}
\left(\begin{array}{cc}
0 & J_p\\
J_p & 0
\end{array}\right)
\mathbf{\hat{\psi}}_{l}+ V\sum\limits_{l=1}^{L} {\hat{n}_{A,l}\hat{n}_{B,l}} 
\end{eqnarray}
with 
$\hat{\psi}_l=(\hat{c}_{A,l},\hat{c}_{B,l})^T$ the annihilation operators of a boson at each sublattice of the $l$th unit cell,
$J_{A,B}^\pm$ and $J_p$ the sublattice-dependent nearest-neighbor hopping and intracell hopping amplitudes respectively,
and $V$ the strength of a intracell interaction. 
In the non-interacting picture with $V=0$, each sublattice represents a copy of the Hatano-Nelson model~\cite{Hatano1996,Hatano1997}, with NHSE induced by asymmetric non-Hermitian hopping amplitudes $J_{A,B}^+\neq J_{A,B}^-$, as shown in~Fig.\ref{fig_Model}(b). 
The localizing direction of skin modes are seen to be reversed by turning on the Hermitian intracell hopping $J_p$,
because it mixes the two sublattices and leads to a destructive interference of their non-reciprocal pumping, resulting in a net non-reciprocity toward the opposite direction~\cite{Li2022PRB}.
It is worth noted that the model we considered here preserves $\mathcal{PT}$ symmetry~\cite{Zhang2022}, with the symmetry operator satisfies ${\cal P}{\cal T}{\hat{\psi}_{l }}{({\cal P}{\cal T})^{ - 1}} = {\hat{\psi}_{L + 1 - l }},{\rm{ }}{\cal P}{\cal T}i{({\cal P}{\cal T})^{ - 1}} =  - i$. 
Under open boundary conditions (OBCs), the single-particle spectrum is real when $J_p=0$ and complex for non-zero $J_p$ (no shown), providing a signature for distinguishing many-body eigenstates with different NHSE in later discussion.
  

\emph{{\clr Occupation-dependent many-body NHSE}.---}
\begin{figure*}
\includegraphics[width=0.95\linewidth]{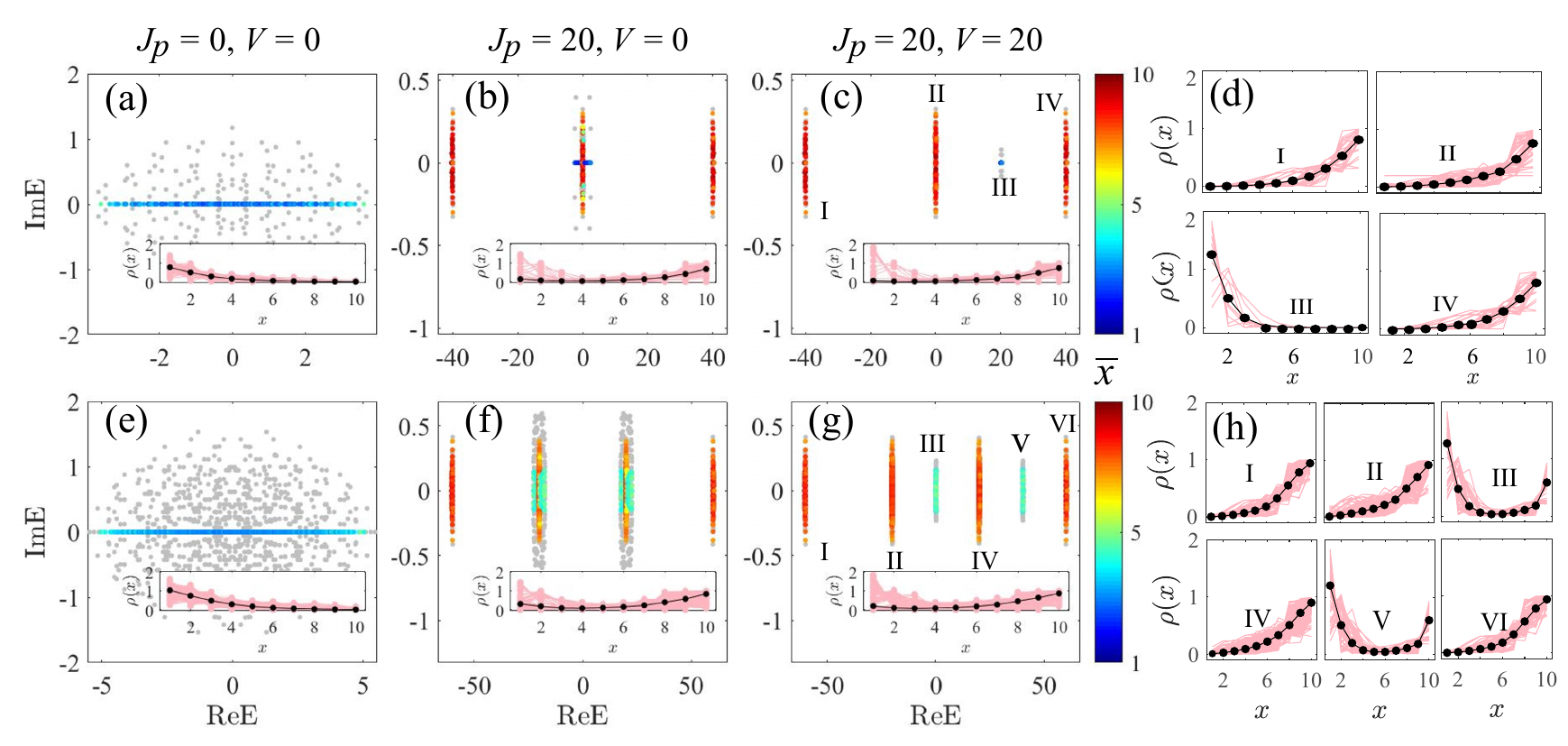}
\caption{Energy spectrum and skin accumulations for $N=2$ and $N=3$. 
For both cases, gray and colored dots indicate energy spectra under PBCs and OBCs respectively, with the colormap representing the mean position of each eigenstate, $\overline{x}=\langle \hat{x}\rangle$.
(a) to (c) Energy spectra for $N=2$ for different values of $J_p$ and $V$, with insets showing the density profiles of all eigenstates (pink) and their average (black).
(d) Density distributions of the four clusters in (c).
Three primary eigenstate clusters with right-NHSE are seen to be induced by the intracell hopping $J_p$, and a secondary one with left-NHSE is separated from them when turning on the interaction $V$.
(e) to (h) the same quantities as in (a) to (d) for our system with $N=3$ particles, which hosts four primary clusters with right-NHSE, and
two secondary clusters exhibiting bipolar-type of skin localization.
$L=10$ is chosen for all panels, other parameters are the same as in Fig.~\ref{fig_Model}.}
\label{fig_N2N3}
\end{figure*}
building upon the hybridization between different sublattices,
the emergence of NHSE and its direction reversal 
are expected to extend
beyond the single-particle picture, 
yet more exotic phenomena may arise 
due to many-body interference.
As shown in Fig. \ref{fig_N2N3}(a) and (b) for $N=2$ particles, all two-body eigenstates accumulate to the left and possess real eigenenergies when intracell hopping is switched off ($J_p=0$), 
but only a part of them has their localizing direction reversed in the presence of strong intracell hopping ($J_p=20$), indicating a co-existence of both types of NHSE in this system for the latter case,
in sharp contrast to the single-particle eigenstates with unidirectional skin localization for the same parameters. 
As further elaborated in Supplemental Materials~\cite{SuppMat}, The coexistence of left- and right-NHSE can be attributed to different occupation of many-body eigenstates, 
giving raise to a spatial separation of paired and unpaired particles occupying the same or different unit cells respectively.
Namely, due to the infinitely large on-site interaction between hardcore bosons, two particles occupying the same unit cell are immune to the intracell hopping, thus manifesting the left-NHSE of the sublattice-decoupled scenario [Fig. \ref{fig_Model}(b)]; while eigenstates with single occupation on each unit cell manifest the right-NHSE of the sublattice-hybridized scenario [Fig. \ref{fig_Model}(c)].

Aside from the distribution profile, a strong intracell hopping is seen to split the
$\mathcal{PT}$-symmetric real eigenenergies in Fig. \ref{fig_N2N3}(a)
into three (primary) clusters in the complex energy plane in Fig. \ref{fig_N2N3}(b), with the real parts of their eigenenergies ${\rm Re}[E]\approx -2J_p,~0,~2J_p$ respectively,
composed by the two single-particle energy bands of the two particles. Note that these clusters are almost ``flat" in their real energy, as a strong intracell hopping induces a destructive interference between the intercell hopping strengths of the two sublattices, resulting in a much smaller net strength.
On top of these, the intracell interaction assigns additional real energy to eigenstates with double occupation, separating a secondary cluster with ${\rm Re}[E]\approx V$ and left-NHSE from the primary ones with right-NHSE, as shown in Fig. \ref{fig_N2N3}(c) and (d).
Consistently, eigenstates in the secondary cluster possesses real eigenenergies as in the sublattice-decoupled scenario.

Similar occupation-dependent NHSE shall also emerge for larger particle number $N$, but is expected to be more sophisticated as 
the system can simultaneously support doubly and singly occupied unit cells.
As shown in Fig. \ref{fig_N2N3}(e) to (h) for $N=3$,
we observe four primary clusters at ${\rm Re}[E]\approx -3J_p,~-J_p,~J_p,~3J_p$ with right-NHSE,
and two secondary clusters at ${\rm Re}[E]\approx\pm J_p+V$ with a bipolar many-body NHSE and complex eigenenergies.
Following the interpretation of occupation-dependency of many-body NHSE, 
when a unit cell is doubly occupied by two particles, the third particle must occupy a different cell and manifests the right-NHSE as in the single-particle level, giving rise to the distinct behaviors of the secondary clusters.
Therefore, the bipolar skin accumulation observed here represents an emergent many-body phenomenon, rather than its single-particle analog characterized by opposite spectral winding numbers for different reference energies \cite{Song2019_PRL}. 
For $N>3$, richer occupation configurations allows for the coexistence of all of the three types of NHSE at different eigenstate clusters. Examples for $N=4$ and $N=5$ are shown in the Supplemental Materials~\cite{SuppMat}.
Note that the left-NHSE emerges only with an even particle number $N$,  so to have a sub-Hilbert space with only doubly occupied states.

\begin{figure}
\centering
\includegraphics[width=0.9\linewidth]{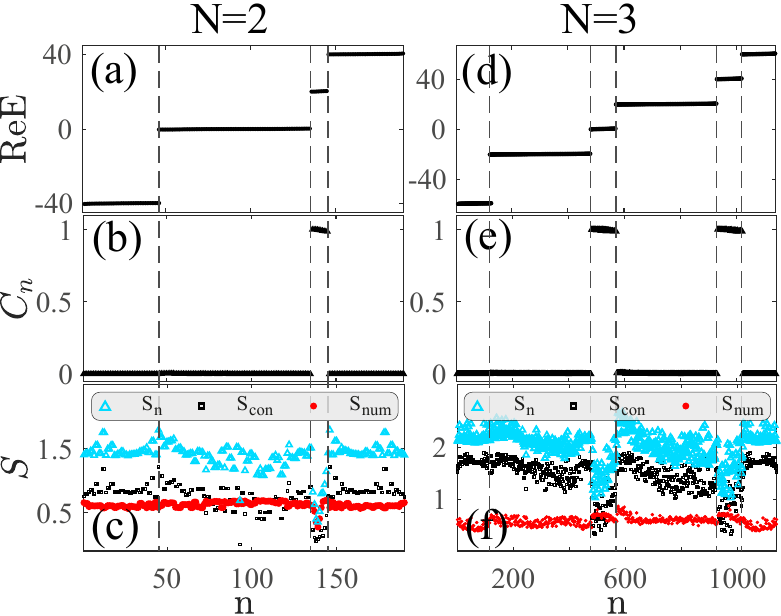}
\caption{\label{fig_S}
Sublattice correlation $C_n$ and entanglement entropy $S_n=S_{\rm num}+S_{\rm con}$, i.e. the sum of particle number entropy and configuration entropy. (a) Real part of the eigenenergy for each eigenstate, (b) sublattice correlation of each eigenstate, and (c) sublattice entanglement entropy, for the system with $N=2$ particles. 
(d-f) the same quantities for $N=3$ with two emerged secondary clusters.
In both cases, emerged secondary clusters with $C_n\approx1$ have smaller $S_n$ compared to that of primary clusters with $C_n\approx0$, serving a signature to distinguish states with 
double and single occupations.
Parameters are $L=10$, $V=20$, and $J_p=20$, with the same $J_{A,B}^{\pm}$ as in Fig.~\ref{fig_Model}(b).
}
\end{figure}

The occupation-dependency of many-body NHSE in our model can be verified by the local sublattice correlation for the each eigenstate, defined as
\begin{eqnarray}
C_n=\sum_{l=1}^{L}\langle \psi_n| \hat{n}_{A,l}\hat{n}_{B,l}| \psi_n\rangle
\end{eqnarray}
with $l$ the cell index and $n$ the index of different eigenstates $|\psi_n\rangle$.
As demonstrated in Fig.~\ref{fig_S} (a, b) and (d, e), primary and secondary clusters have their eigenstates with $C_n\approx0$ and $1$, indicating their 
single- and double-occupation of particles
respectively.
To further characterize this phenomenon, we calculate the entanglement entropy between two sublattices, defined as
\begin{eqnarray}
S_{n}=-{\rm Tr} \rho_{n,A}\ln \rho_{n,A}=S_{\rm num}+S_{\rm con},
\end{eqnarray}
with $\rho_{n,A}={\rm Tr}_{B} [\rho_n]={\rm Tr}_{B}[| \psi_n\rangle \langle \psi_n|]$, and
$S_{\rm num}$ and $S_{\rm con}$ the particle number entropy and configuration entropy respectively~\cite{lukin2019probing,Orito2022},
\begin{equation}
\begin{array}{l}
{S_{{\rm{num}}}} =  - \sum\limits_{{N_A}} {{p_{{N_A}}}\ln {p_{{N_A}}},} \\
{S_{{\rm{con}}}} =  - \sum\limits_{{N_A}} {\sum\limits_\alpha  {{p_{{N_A}}}\tilde \lambda _\alpha ^{\left( {{N_A}} \right)}\ln } } \tilde \lambda _\alpha ^{\left( {{N_A}} \right)}.
\end{array}
\end{equation}
 Here we drop the eigenstate index `$n$' for simplicity.
 It is worth to mention that $N_{A}$ and $\rho_{A}$ are simultaneously diagonalizable, 
so that 
${\rho _{A}} = {\rho _{{N_1} \oplus }}{\rho _{{N_2} \oplus }}{\rho _{{N_3} \oplus }}...$,
and ${\lambda _\alpha ^{\left( {{N_l}} \right)}}$ is the eigenvalues of $\rho_{N_{l}}$ with $l=1,2,3,...$ and $N_l\in\{0,1,...,N\}$. ${{p_{{N_A}}}}$ and ${\tilde \lambda _\alpha ^{\left( {{N_A}} \right)}}$ are defined as 
\begin{equation}
{p_{{N_A}}} = \sum\limits_\alpha  {\lambda _\alpha ^{\left( {{N_A}} \right)}} ,\tilde \lambda _\alpha ^{\left( {{N_A}} \right)} = \frac{1}{{{p_{{N_A}}}}}\lambda _\alpha ^{\left( {{N_A}} \right)}.
\end{equation}
The number entropy quantifies the number fluctuations in the subsystem, while the configuration entropy quantifies the nonlocal correlations.
As observed in Fig.~\ref{fig_S} (c) and (f), 
$S_{n}$ takes smaller values for secondary clusters with 
doubly occupied states,
as they have weaker sublattice fluctuation and less state-configurations in the subsystem of one sublattice.

\emph{{\clr Dynamical separation of paired and unpaired particles}.---} 
The
 occupation-dependency of many-body NHSE can also be expected to manifest during the dynamics of states with different configurations of
single and double occupations.
Specifically, we consider an initial Fock state $|\psi(0)\rangle$ with 
hardcore bosons localized at the center of the system,
and normalize its evolved state at time $t$ as
\begin{equation}
\left|\psi(t)\right\rangle=\frac{e^{-i \hat{H} t / \hbar}\left|\psi_0\right\rangle}{\sqrt{\left\langle\psi_0\left|e^{i \hat{H}^{\dagger} t / \hbar} e^{-i \hat{H} t / \hbar}\right| \psi_0\right\rangle}}.
\end{equation}
In Fig.~\ref{fig_D} (a) to (d),
we displayed the particle density 
$\rho(x,t)=\langle\psi(t)| \hat{n}(x)\left|\psi(t)\right\rangle$
for initial states with different double or single occupation.
As clearly shown in Fig.~\ref{fig_D} (a) and (b),  
paired particles initially occupying the same unit cell move to the left-hand side for $N=2$ with initial state $| (AB)_5\rangle$ and $N=3$ with initial state $| (AB)_4A_5\rangle$, respectively,
and accumulate at the boundary within our observation time $t \in (0,20)$.
It is worth to mention that due to the large imaginary energy of the primary clusters, the evolved state will eventually localize at the right side after a sufficient long time, which also lead to distinguished short-time and long-time dynamics of entanglement entropy~\cite{SuppMat}. 
Meanwhile, 
the unpaired particle for $N=3$ (centered at the fifth lattice site in our example) is seen to move to the right, manifesting the bipolar-NHSE together with the paired ones. 
On the other hand, initial states with only unpaired particles, e.g. $|B_4B_5\rangle$ and 
$|B_3B_4B_5\rangle$ for $N=2$ and $N=3$ respectively, are governed by the right-NHSE and move to the right-hand side, as shown in Fig.~\ref{fig_D} (c) and (d). 
Note that these states become less localized during the evolution, because
(i) the initial states are already relatively extended compared with the system's size ($2$ or $3$ sites for $L=8$);
and (ii) the right-NHSE is considerably weaker as it origins from destructive interference of non-reciprocity [as can be seen from the two branches of time-evolving modes in Fig.~\ref{fig_D}(b)].
\begin{figure}
\centering
\includegraphics[width=\linewidth]{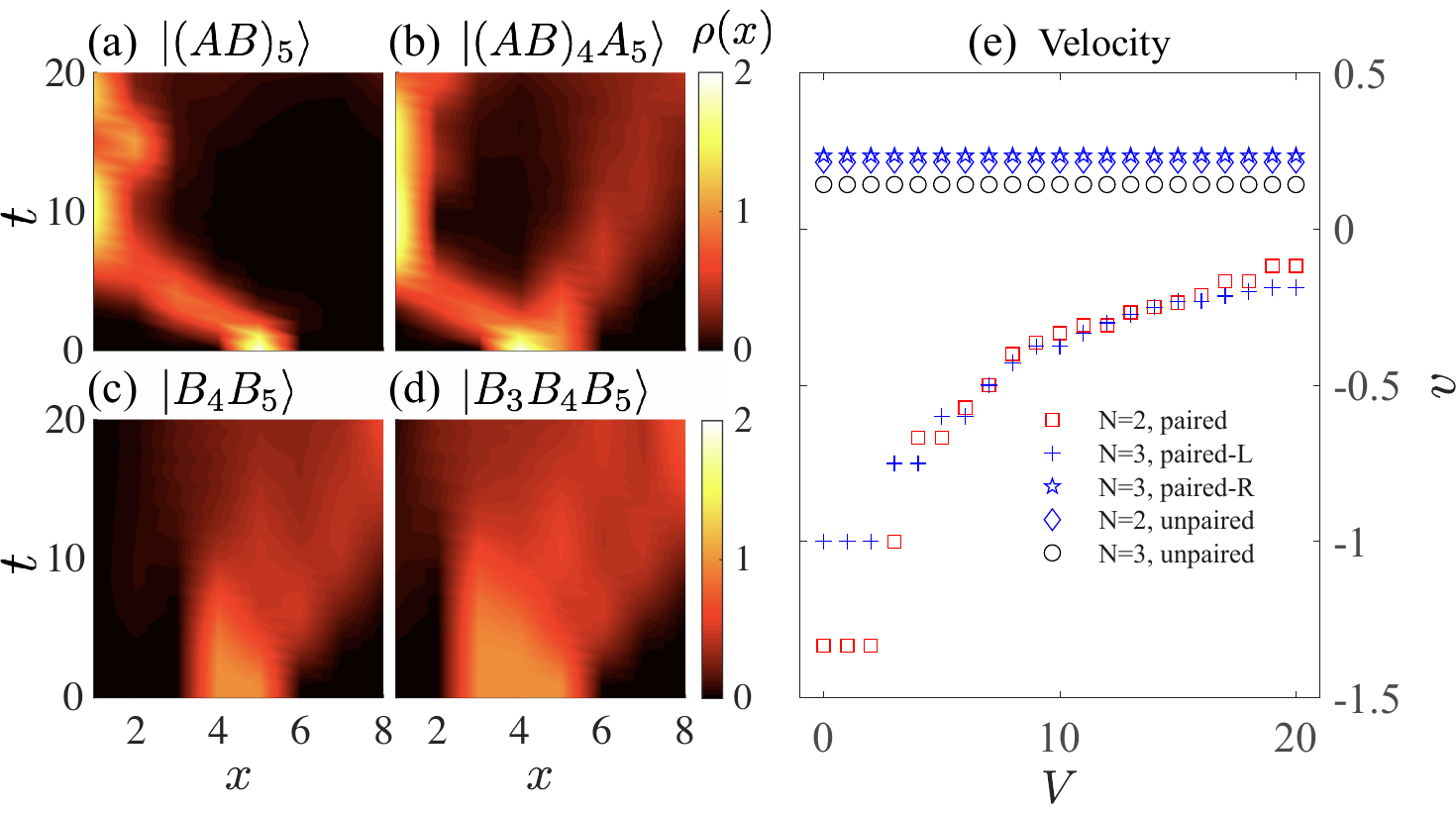}
\caption{\label{fig_D}
Dynamics evolution for initial states with different single and double occupation.
(a) to (d) illustrate the density profiles of the evolved state $|\psi(t)\rangle$, for different initial states $|\psi(0)\rangle$
given by
(a) $|(AB)_5\rangle$, 
(b) $|(AB)_4A_5\rangle$, 
(c) $|B_4B_5\rangle$, and 
(d) $|B_3B_4B_5\rangle$,
with $J_p=20$, $V=5$ and $L=8$.
Paired and unpaired particles are seen to move to opposite directions.
(e) The velocity of density peak for (a-d) versus interaction $V$. 
Red squares (blue crosses) refer to the left-moving peak of $N=2$ ($N=3$) with doubly occupied initial state. Blue rhombuses (black circles) refer to the right-moving peak of $N=2$ ($N=3$) with singly occupied initial state. Blue stars refer to the right-moving peak of $N=3$ with doubly occupied initial state. As can be seen, the velocity is almost unchanged for singly occupied initial states of  $|B_4B_5\rangle$ and $|B_3B_4B_5\rangle$ (or the unpair particle in $|(AB)_4A_5\rangle$) with different $V$, as these states are immune to the intracell interaction. 
Other parameters no mentioned here are the same as in Fig.~\ref{fig_Model}.}
\end{figure}

Finally, to quantify the interaction effects on the dynamics, we define the velocity of peak density for both directions as
\begin{equation}
v_p=\frac{x_p(t)-x_p(0)}{t},
\end{equation}
where $x_p(0)$ ($x_p(t)$) is the positions with maximal density of $|\psi(0)\rangle$ ($|\psi(t)\rangle$).
As an exception, 
for the double occupied initial state $|(AB)_4A_5\rangle$ with $N=3$,
we consider two local maximal values with $x_p^+(t)>x_p(0)$ and $x_p^-(t)<x_p(0)$,
to characterize the two separated branches of time-evolving modes in Fig.~\ref{fig_D}(b) respectively.
Numerically, we choose $t$ to be the time for the density peak $x_p(t)$ to reach the left or right boundary.
As observed in Fig.~\ref{fig_D}(e), 
unpaired particles have almost identical velocity as in the non-interacting limit, as they are not affected by the intracell interaction we consider.
In contrast, the absolute value of the left moving velocity for paired particles decreases with increasing $V$, for both $N=2$ and $N=3$. Following our perturbation calculation for $N=2$~\cite{SuppMat}, effective hopping amplitudes of paired particles are proportional to $1/V$,
thus leading to the weaker skin accumulation at larger $V$ (with numerical results demonstrated in Supplemental Materials~\cite{SuppMat}).



\par
\emph{{\clr Summary and conclusions}.---}
We have shown that the many-body effects of hardcore bosons can induce distinct skin clusters in systems with multiple non-reciprocal pumping channels.
The resultant emergent
occupation-dependency 
of many-body NHSE is justified with both perturbation calculation for a sub-Hilbert space of doubly occupied basis states, and numerical results of sublattice correlation and inter-sublattice entanglement entropy.
In the minimal model we present, the emergent skin clusters also shows a dependence on the parity of particle number, 
as different parities allow for different particle configurations.
Namely, a system with an odd number of particles support only right- and bipolar-NHSE, and left-NHSE emerges only for an even number of particles.
Finally, we show that during a time evolution, the 
occupation-dependency of NHSE manifest as a dynamical separation of paired and unpaired particles.
We note that in a different parameter regime, intracell hopping in our model may lead to vanishing NHSE even in the presence of asymmetric hoppings, which results in another type of occupation-dependent dynamics, as demonstrated in Supplemental Materials~\cite{SuppMat}.

The highlight of this work is the selective activation of different NHSE channels for paired and unpaired particles, which 
not only provides an efficient mechanism for separating particles accordingly to their occupation configurations, but also
opens an avenue toward exploring emergent many-body non-Hermitian effects built on non-conservation and configurations of particles in subsystems, such as sublattices, orbitals, and spin species.
As illustrate in Supplemental Materials~\cite{SuppMat}, a similar particle separation also occurs for spin-1/2 fermions, where paired and unpaired spins can be separated in the presence of strong spin flipping and interactions.
 These occupation/pairing-dependent NHSE and dynamics cannot be simply deduced from single-particle behaviors, where all eigenstates
are skin modes localized at the same end of the system.
Originated from the many-body nature, our found phenomena are mostly suitable to be implemented in cold atoms, where NHSE has been predicted and realized in various real- or momentum-space lattices~\cite{li2020topological,liang2022dynamic,guo2022theoretical,li2022BEC,zhou2022engineering,yi2023geometry}.
For instance, 
realization of coupled chains with different non-reciprocal pumping directions have been theoretically proposed in an optically shaken lattice accommodating cold atoms~\cite{li2020topological}, where the two chains may be mapped to the two sublattices in our system. 
Following the framework we provide, more exciting phenomena may be found in systems with
long-range interaction, which can generate richer particle configurations (e.g. the density wave states);
or higher spatial dimensions, which support more sophisticated non-reciprocal pumping channels on different bulk and boundaries. 
Disorders and defects may also lead to enigmatic many-body NHSE based on particle configurations, as they affect both allowed spatial configurations and non-reciprocity of the system.

\emph{{\clr Acknowledgements}.---}
L. Li would like to thank C. H. Lee for helpful discussion.
This work is
supported by National Natural Science Foundation of China (Grant No. 12104519) and the
Guangdong Project (Grant No. 2021QN02X073).

\newpage

\begin{widetext}

	\setcounter{equation}{0} \setcounter{figure}{0} \setcounter{table}{0} %
	\renewcommand{\theequation}{S\arabic{equation}} \renewcommand{\thefigure}{S%
		\arabic{figure}} \renewcommand{\bibnumfmt}[1]{[S#1]} 
	\renewcommand{\citenumfont}[1]{S#1}


\begin{center}
\textbf{\large Supplementary Materials for ``Occupation-dependent particle separation in one-dimensional non-Hermitian lattices"}
\end{center}


\section{Perturbation treatment for $N=2$ }\
In this section we apply a perturbation treatment to analyze the non-Hermitian skin effect (NHSE) for
states with $N=2$ paired particles occupying the same unit cell. Intuitively, we shall obtain an effective model in a $L$-dimensional sub-Hilbert space (with $L$ the number of unit cells), 
which excludes intracell hopping.
Explicitly, we define basis in this subspace as 
\begin{equation}
\left. {\left| {{\alpha _i}} \right\rangle } \right\rangle  = \hat{c}_{A,i}^\dag \hat{c}_{B,i}^\dag \left| {{\rm{vac}}} \right\rangle  \equiv \left| {{{\left( { A B} \right)}_i}} \right\rangle {\rm{,   (}}i = 1,2,...,L{\rm{)}}{\rm{,}}
\end{equation}
with $i=1,...,L$ and $\left| {{\rm{vac}}} \right\rangle$ the vacuum state. The Hamiltonian can be expanded as 
\begin{equation}
\hat{H}_{\rm eff}=\sum_{i,j}\langle\langle \alpha_j|\hat{H}| \alpha_i \rangle\rangle  | \alpha_j \rangle\rangle \langle\langle  \alpha_i|.
\end{equation}
Thus, the intracell hopping term $\langle\langle \alpha_j|J_p\sum_l(\hat{c}^{\dagger}_{A, l}\hat{c}_{B, l}+\hat{c}^{\dagger}_{B, l}\hat{c}_{A,l})| \alpha_i \rangle\rangle$ should always be zero. 
Next, we use second order perturbation theory to deal with the two particles case in the 
doubly occupied subspace. 
For the tight-binding Hamiltonian 
\begin{equation}
\begin{array}{*{20}{l}}
\hat{H}{ = \sum\limits_{l = 1}^{L - 1} {\left[ {J_ A ^ +\hat{c}_{ A ,l + 1}^\dag {\hat{c}_{A,l}} + J_A^ - \hat{c}_{A,l}^\dag {\hat{c}_{ A ,l + 1}}} \right]}  + \sum\limits_{l = 1}^{L - 1} {\left[ {J_B^ + \hat{c}_{ B,l + 1}^\dag {\hat{c}_{ B ,l}} + J_B^ - \hat{c}_{B ,l}^\dag {\hat{c}_{ B ,l + 1}}} \right]} }{ + \sum\limits_{l = 1}^L {J_p^{}\left[ {\hat{c}_{ B ,l}^\dag {\hat{c}_{ A ,l}} + \hat{c}_{ A ,l}^\dag {\hat{c}_{ B ,l}}} \right]}  + V\sum\limits_{l = 1}^L {\hat{n}_{B,l}\hat{n}_{A,l} ,} }
\end{array}
\end{equation}
we take the interaction term as unperturbed Hamiltonian,
\[{\hat H_{\rm int}} = V\sum\limits_{l = 1}^L {\hat{n}_{B,l} \hat{n}_{A,l}} ,\]
and the rest hopping terms as perturbation,
\[\begin{array}{*{20}{l}}
\hat{H}_{\rm hop}{ = \sum\limits_{l = 1}^{L - 1} {\left[ {J_ A ^ + \hat{c}_{A ,l + 1}^\dag {\hat{c}_{A,l}} + J_A^ - \hat{c}_{A,l }^\dag {\hat{c}_{ A ,l + 1}}} \right]}  + \sum\limits_{l = 1}^{L - 1} {\left[ {J_B^ + \hat{c}_{ B ,l + 1}^\dag {\hat{c}_{ B ,l}} + J_B^ - \hat{c}_{ B ,l}^\dag {\hat{c}_{ B ,l + 1}}} \right]} }{ + \sum\limits_{l = 1}^L {J_p^{}\left[ {\hat{c}_{ B ,l}^\dag {\hat{c}_{ A ,l}} + \hat{c}_{A,l}^\dag {\hat{c}_{ B ,l}}} \right]}  . }
\end{array}\]

For $N=2$, the unperturbed eigenenergy is always $E=V$ in the 
paired subspace,
and the effective Hamiltonian ${{\hat{H}}_{\text{eff}}}$ is obtained as 
$${\hat H_{{\rm{eff}}}} = E + {\hat{{\cal P}}_{{\rm{int}}}}{\hat H_{{\rm{hop}}}}{\hat{{\cal P}}_{{\rm{int}}}} + {\hat{{\cal P}}}_{{\rm{int}}}{\hat H_{{\rm{hop}}}}{\left( {E - {{\hat H}_{{\rm{int}}}}} \right)^{ - 1}}{\hat H_{{\rm{hop}}}}{\hat{{\cal P}}_{{\rm{int}}}} + {\cal O}\left( {\hat H_{{\rm{hop}}}^3} \right),$$
with
$\hat{{\cal P}}_{{\rm{int}}} = \sum\nolimits_{i = 1}^L {\left| {\left. {{\alpha _i}} \right\rangle } \right\rangle } \left\langle {\left\langle {{\alpha _i}} \right.} \right|$ the projector onto the 
doubly occupied sub-Hilbert space. 
The first-order perturbation ${\hat{{\cal P}}_{{\rm{int}}}}{\hat {H}_{\rm hop}}{\hat{{\cal P}}_{{\rm{int}}}}$ shall be zero, 
since ${\hat {H}_{\rm hop}}$  always split two paired particles to two adjacent sites.
The matrix elements of the second-order correction term are
\begin{eqnarray}
\left\langle {\left\langle {{\alpha _j}} \right.} \right|{{\hat H}_{{\rm{hop}}}}{\left( {E - {{\hat H}_{{\rm{int}}}}} \right)^{ - 1}}{{\hat H}_{{\rm{hop}}}}\left| {\left. {{\alpha _i}} \right\rangle } \right\rangle 
 = \left\langle {{\rm{vac}}} \right|\hat{c}_{B,j} \hat{c}_{A,j} {{\hat H}_{{\rm{hop}}}}{\left( {E - {{\hat H}_{{\rm{int}}}}} \right)^{ - 1}}{{\hat H}_{{\rm{hop}}}}\hat{c}_{A,i}^\dag \hat{c}_{B,i}^\dag \left| {{\rm{vac}}} \right\rangle,\label{Seq:elemant}
\end{eqnarray}
where
\[\begin{array}{llll}
&{{\hat H}_{{\rm{hop}}}}{\left( {E - {{\hat H}_{{\rm{int}}}}} \right)^{ - 1}}{{\hat H}_{{\rm{hop}}}} = {{\hat H}_{{\rm{hop}}}}{\left( {E - V\sum\limits_{l = 1}^L {\hat{n}_{A,l} \hat{n}_{B,l} } } \right)^{ - 1}}{{\hat H}_{{\rm{hop}}}}\\
 & = \left(  \sum\limits_{l = 1}^{L - 1} {\left[ {J_ A ^ + \hat{c}_{ A ,l + 1}^\dag {\hat{c}_{ A ,l}} + J_ A ^ - \hat{c}_{ A ,l}^\dag {\hat{c}_{ A ,l + 1}}} \right]}  + \sum\limits_{l = 1}^{L - 1} {\left[ {J_ B ^ + \hat{c}_{ B ,l + 1}^\dag {\hat{c}_{ B ,l}} + J_ B ^ - \hat{c}_{ B ,l}^\dag {\hat{c}_{ B ,l + 1}}} \right] + \sum\limits_{l = 1}^L {J_p^{}\left[ {\hat{c}_{ B ,l}^\dag {\hat{c}_{ A ,l}} + \hat{c}_{ A ,l}^\dag {\hat{c}_{ B ,l}}} \right]} }  \right)\times\\
 &{\left( {E - V\sum\limits_{l = 1}^L {n_l^ B n_l^ A } } \right)^{ - 1}} \times \\
&\left( {  \sum\limits_{l = 1}^{L - 1} {\left[ {J_ A ^ + \hat{c}_{ A ,l + 1}^\dag {\hat{c}_{ A ,l}} + J_ A ^ - \hat{c}_{ A ,l}^\dag {\hat{c}_{ A ,l + 1}}} \right]}  + \sum\limits_{l = 1}^{L - 1} {\left[ {J_ B ^ + \hat{c}_{ B ,l + 1}^\dag {\hat{c}_{ B ,l}} + J_ B ^ - \hat{c}_{ B ,l}^\dag {\hat{c}_{ B ,l + 1}}} \right]}  + \sum\limits_{l = 1}^L {J_p^{}\left[ {\hat{c}_{ B ,l}^\dag {\hat{c}_{ A ,l}} + \hat{c}_{ A ,l}^\dag {\hat{c}_{ B ,l}}} \right]} } \right).
\end{array}\]
Applying each operator to $| \alpha _i \rangle \rangle  = \hat{c}_{A,i}^\dag \hat{c}_{B,i}^\dag | {\rm vac} \rangle$ in Eq.~\eqref{Seq:elemant}, we obtain

\[\begin{array}{l}
\left( {  \sum\limits_{l = 1}^{L - 1} {\left[ {J_ A ^ + \hat{c}_{ A ,l + 1}^\dag {\hat{c}_{ A ,l}} + J_ A ^ - \hat{c}_{ A ,l}^\dag {\hat{c}_{ A l + 1}}} \right]}  + \sum\limits_{l = 1}^{L - 1} {\left[ {J_ B ^ + \hat{c}_{ B ,l + 1}^\dag {\hat{c}_{ B ,l}} + J_ B ^ - \hat{c}_{ B ,l}^\dag {\hat{c}_{ B ,l + 1}} + } \right] + \sum\limits_{l = 1}^L {J_p^{}\left[ {\hat{c}_{ B ,l}^\dag {\hat{c}_{ A ,l}} + \hat{c}_{ A ,l}^\dag {\hat{c}_{ B ,l}}} \right]} } } \right)\hat{c}_{ A ,i}^\dag \hat{c}_{ B ,i}^\dag \left| {{\rm{vac}}} \right\rangle \\
 = \left( {J_ A ^ + \hat{c}_{ A ,i + 1}^\dag \hat{c}_{ B ,i}^\dag  + J_ A ^ - \hat{c}_{ A ,i - 1}^\dag \hat{c}_{ B ,i}^\dag  + J_ B ^ + \hat{c}_{ A ,i}^\dag \hat{c}_{ B ,i + 1}^\dag  + J_ B ^ - \hat{c}_{ A ,i}^\dag \hat{c}_{ B ,i - 1}^\dag } \right)\left| {{\rm{vac}}} \right\rangle;
\end{array}\]
\[\begin{array}{l}
{\left( {E - V\sum\limits_{l = 1}^L {\hat{n}_{A,l} \hat{n}_{B,l} } } \right)^{ - 1}}\left( {J_ A ^ + \hat{c}_{ A ,i + 1}^\dag \hat{c}_{ B ,i}^\dag  + J_ A ^ - \hat{c}_{ A ,i - 1}^\dag \hat{c}_{ B ,i}^\dag  + J_ B ^ + \hat{c}_{ A ,i}^\dag \hat{c}_{ B ,i + 1}^\dag  + J_ B ^ - \hat{c}_{ A ,i}^\dag \hat{c}_{ B ,i - 1}^\dag } \right)\left| {{\rm{vac}}} \right\rangle \\
 = \frac{1}{V}\left( {J_ A ^ + \hat{c}_{ A ,i + 1}^\dag \hat{c}_{ B ,i}^\dag  + J_ A ^ - \hat{c}_{ A ,i - 1}^\dag \hat{c}_{ B ,i}^\dag  + J_ B ^ + \hat{c}_{ A ,i}^\dag \hat{c}_{ B ,i + 1}^\dag  + J_ B ^ - \hat{c}_{ A ,i}^\dag \hat{c}_{ B ,i - 1}^\dag } \right)\left| {{\rm{vac}}} \right\rangle; \\
 \end{array}\]
 \[\begin{array}{l}
\left( {  \sum\limits_{l = 1}^{L - 1} {\left[ {J_ A ^ + \hat{c}_{ A ,l + 1}^\dag {\hat{c}_{ A ,l}} + J_ A ^ - \hat{c}_{ A ,l}^\dag {\hat{c}_{ A ,l + 1}}} \right]}  + \sum\limits_{l = 1}^{L - 1} {\left[ {J_ B ^ + \hat{c}_{ B ,l + 1}^\dag {\hat{c}_{ B ,l}} + J_ B ^ - \hat{c}_{ B ,l}^\dag {\hat{c}_{ B ,l + 1}}} \right]}  + \sum\limits_{l = 1}^L {{J_p}\left[ {\hat{c}_{ B ,l}^\dag {\hat{c}_{ A ,l}} + \hat{c}_{ A ,l}^\dag {\hat{c}_{ B ,l}}} \right]} } \right) \times \\
\frac{1}{V}\left( { + J_ A ^ + \hat{c}_{ A ,i + 1}^\dag \hat{c}_{ B ,i}^\dag  + J_ A ^ - \hat{c}_{ A ,i - 1}^\dag \hat{c}_{ B ,i}^\dag  + J_ B ^ + \hat{c}_{ A ,i}^\dag \hat{c}_{ B ,i + 1}^\dag  + J_ B ^ - \hat{c}_{ A ,i}^\dag \hat{c}_{ B ,i - 1}^\dag } \right)\left| {{\rm{vac}}} \right\rangle \\
 = \frac{1}{V}\left( {J_ A ^ + J_ A ^ + \hat{c}_{ A ,i + 2}^\dag \hat{c}_{ B ,i}^\dag  + J_ A ^ + J_ A ^ - \hat{c}_{ A ,i}^\dag \hat{c}_{ B ,i}^\dag  + J_ A ^ + J_ B ^ + \hat{c}_{ A ,i + 1}^\dag \hat{c}_{ B ,i + 1}^\dag  + J_ A ^ + J_ B ^ - \hat{c}_{ A ,i + 1}^\dag \hat{c}_{ B ,i - 1}^\dag } \right)\left| {{\rm{vac}}} \right\rangle \\
 + \frac{1}{V}\left( {J_ A ^ - J_ A ^ + \hat{c}_{ A ,i}^\dag \hat{c}_{ B ,i}^\dag  + J_ A ^ - J_ A ^ - \hat{c}_{ A ,i - 2}^\dag \hat{c}_{ B ,i}^\dag  + J_ A ^ - J_ B ^ + \hat{c}_{ A ,i - 1}^\dag \hat{c}_{ B ,i + 1}^\dag  + J_ A ^ - J_ B ^ - \hat{c}_{ A ,i - 1}^\dag \hat{c}_{ B ,i - 1}^\dag } \right)\left| {{\rm{vac}}} \right\rangle \\
 + \frac{1}{V}\left( {J_ B ^ + J_ A ^ + \hat{c}_{ A ,i + 1}^\dag \hat{c}_{ B ,i + 1}^\dag  + J_ B ^ + J_ A ^ - \hat{c}_{ A ,i - 1}^\dag \hat{c}_{ B ,i + 1}^\dag  + J_ B ^ + J_ B ^ + \hat{c}_{ A ,i}^\dag \hat{c}_{ B ,i + 2}^\dag  + J_ B ^ + J_ B ^ - \hat{c}_{ A ,i}^\dag \hat{c}_{ B ,i}^\dag } \right)\left| {{\rm{vac}}} \right\rangle \\
+ \frac{1}{V}\left( {J_ B ^ - J_ A ^ + \hat{c}_{ A ,i + 1}^\dag \hat{c}_{ B ,i - 1}^\dag  + J_ B ^ - J_ A ^ - \hat{c}_{ A ,i - 1}^\dag \hat{c}_{ B ,i - 1}^\dag  + J_ B ^ - J_ B ^ + \hat{c}_{ A ,i}^\dag \hat{c}_{ B ,i}^\dag  + J_ B ^ - J_ B ^ - \hat{c}_{ A ,i}^\dag \hat{c}_{ B ,i - 2}^\dag } \right)\left| {{\rm{vac}}} \right\rangle.
\end{array}\]

Finally, we obtain the matrix element in $\hat{H}_{\rm eff}$ as
\[\begin{array}{*{20}{l}}
{\langle {\rm{vac}}|{\hat{c}_{ A ,j}}{\hat{c}_{ B ,j}}{{\hat H}_{{\rm{hop}}}}{{\left( {E - {{\hat H}_{{\rm{int}}}}} \right)}^{ - 1}}{{\hat H}_{{\rm{hop}}}}\hat{c}_{ A ,i}^\dag \hat{c}_{ B ,i}^\dag \left| {{\rm{vac}}} \right\rangle }\\
{ = \frac{1}{V}\left( {J_ A ^ + J_ A ^ - {\delta _{j,i}} + J_ B ^ + J_ A ^ + {\delta _{j,i + 1}} + J_ A ^ - J_ A ^ + {\delta _{j,i}} + J_ A ^ - J_ B ^ - {\delta _{j,i - 1}} + }  {J_ B ^ + J_ A ^ + {\delta _{j,i + 1}} + J_ B ^ + J_ B ^ - {\delta _{i,j}} + J_ B ^ - J_ A ^ - {\delta _{j,i - 1}} + J_ B ^ - J_ B ^ + {\delta _{j,i}}} \right)}\\
{ = \frac{2}{V}\left( {J_ A ^ + J_ A ^ -  + J_ B ^ + J_ B ^ - } \right){\delta _{ji}} + \frac{2}{V}J_ A ^ - J_ B ^ - {\delta _{j,i - 1}} + \frac{2}{V}J_ B ^ + J_ A ^ + {\delta _{j,i + 1}}.}
\end{array}\]
The effective Hamiltonian is 
\[\begin{array}{*{20}{l}}
{{{\hat H}_{{\rm{eff}}}} \simeq V + \frac{2}{V}\left( {J_ A ^ + J_ A ^ -  + J_ B ^ + J_ B ^ - } \right) + }
{\sum\limits_{i = 1}^L {\left[ {\frac{2}{V}J_ B ^ + J_ A ^ + \left|  {{\alpha _{i + 1}}} \right. \rangle \rangle \langle\langle \left. {{\alpha _i}} \right| + \frac{2}{V}J_ A ^ - J_ B ^ - \left|  {{\alpha _i}}  \right\rangle\rangle \rangle \langle  {{\alpha _{i + 1}}}  |} \right]} ,}
\end{array}\]
i.e. the Hatano-Nelson model \cite{Hatano1996S,Hatano1997S} with asymmetric hopping amplitudes $J_{A}^{\pm}J_{B}^{\pm}$.
With a Fourier  transformation ($i\rightarrow \theta$), 
its eigenenergies are given by

\[\begin{array}{*{20}{l}}
E&{ \simeq V + \frac{2}{V}\left( {J_ A ^ + J_ A ^ -  + J_ B ^ + J_ B ^ - } \right) + \frac{2}{V}J_ B ^ + J_ A ^ + {e^{ - i\theta }} + \frac{2}{V}J_ A ^ - J_ B ^ - {e^{i\theta }}}\\
{}&{ = V + \frac{2}{V}\left( {J_ A ^ + J_ A ^ -  + J_ B ^ + J_ B ^ - } \right) + \frac{2}{V}\left( {J_ B ^ + J_ A ^ +  + J_ A ^ - J_ B ^ - } \right)\cos \theta }{ + i\frac{2}{V}\left( {J_ A ^ - J_ B ^ -  - J_ B ^ + J_ A ^ + } \right)\sin \theta ,}
\end{array}\]

with $\theta\in[0,2\pi)$.
Its energy spectrum forms a loop in the complex energy plane when  \[J_{B}^{+}J_{A}^{+}\ne J_{A}^{-}J_{B}^{-},\] where NHSE shall occur under open boundary condition (OBC).
When \[|J_{A}^{-}J_{B}^{-}|>|J_{A}^{+}J_{B}^{+}|,\] the OBC eigenstates are localized at the left edge of the system. Note that in the above derivations, $V\gg J_p$ shall be assumed to apply the perturbation treatment,
yet numerically we observe left localization of eigenstates in the secondary cluster even for $V\approx J_p$.

\section{Exact solution for $N=3$ particles in two unit cells}\
To gain some insights on the bipolar-NHSE of $N=3$ case, we consider the simplest case with $L=2$ and $N=3$. The Hamiltonian reads 

\begin{equation}
\begin{array}{*{20}{c}}
{{\hat{H}_{{\rm{2 - site}}}}}&{ = \sum\limits_{i = 1}^2 {\left[ {{J_p}\left( {\hat{c}_{ A ,i}^\dag {\hat{c}_{ B ,i}} + \hat{c}_{ B ,i}^\dag {\hat{c}_{ A ,i}}} \right) + V\hat{c}_{ A ,i}^\dag {\hat{c}_{ A ,i}}\hat{c}_{ B ,i}^\dag {\hat{c}_{ B ,i}}} \right]} }\\
{}&{ + J_ A ^ - \hat{c}_{ A ,1}^\dag {\hat{c}_{ A ,2}} + J_ A ^ + \hat{c}_{ A ,2}^\dag {\hat{c}_{ A ,1}} + J_ B ^ - \hat{c}_{ B ,1}^\dag {\hat{c}_{ B ,2}} + J_ B ^ + \hat{c}_{ B ,2}^\dag {\hat{c}_{ B ,1}}.}
\end{array}
\end{equation}
The basis in Fock space is 
\begin{equation}
\begin{array}{l}
\left| {{\phi _1}} \right\rangle  = \left| {{{( A  B )}_1}{ A _2}} \right\rangle ,\left| {{\phi _2}} \right\rangle  = \left| {{{( A  B )}_1}{ B _2}} \right\rangle ,\\
\left| {{\phi _3}} \right\rangle  = \left| {{ A _1}{{( A  B )}_2}} \right\rangle ,\left| {{\phi _4}} \right\rangle  = \left| {{ B _1}{{( A  B )}_2}} \right\rangle ,
\end{array}
\end{equation}
where the subscript index of $A$ and $B$ labels the number of each unit cell. 
The matrix form of Hamiltonian in this basis is given by  

\begin{equation}
{H_{{\rm{2 - site}}}} = \left( {\begin{array}{*{20}{c}}
{\begin{array}{*{20}{c}}
V&{{J_p}}\\
{{J_p}}&V
\end{array}}&{\begin{array}{*{20}{c}}
{J_{^B}^ +}&0\\
0&{J_{^A}^ + }
\end{array}}\\
{\begin{array}{*{20}{c}}
{J_{^B}^ - }&0\\
0&{J_{^A}^ - }
\end{array}}&{\begin{array}{*{20}{c}}
V&{{J_p}}\\
{{J_p}}&V
\end{array}}
\end{array}} \right).
\end{equation}
The eigenvalues are 
\begin{equation}
\begin{array}{l}
{E_1} = V - \sqrt {\frac{{f - g}}{2}} ,{E_2} = V + \sqrt {\frac{{f - g}}{2}} ,\\
{E_3} = V - \sqrt {\frac{{f + g}}{2}} ,{E_4} = V + \sqrt {\frac{{f + g}}{2}} ,
\end{array}
\end{equation}
with $f$ and $g$ defined as

\begin{equation}
\begin{array}{l}
f \equiv 2J_p^2 + J_A^ - J_A^ +  + J_B^ - J_B^ + ,{\rm{ }}\\
g = \sqrt {{{\left( {J_A^ - J_A^ +  - J_B^ - J_B^ + } \right)}^2} + 4\left( {J_A^ -  + J_B^ - } \right)\left( {J_A^ +  + J_B^ + } \right)J_p^2}. 
\end{array}
\end{equation}
The eigenvectors (of the matrix) are

\[\begin{array}{*{20}{l}}
{{\varphi _1} = \left\{ { - \frac{{h + g}}{{2(J_ A ^ -  + J_ B ^ - ){J_p}}},\sqrt {\frac{{f - g}}{2}} \frac{{h - g}}{{\left( {g - hJ_ B ^ -  - 2(J_ A ^ -  + J_ B ^ - )J_p^2} \right)}}} \right.,\left. {\frac{{\sqrt 2 (J_ A ^ -  + J_ B ^ - ){J_p}\sqrt {f - g} }}{{\left( {g - hJ_ A ^ -  - 2(J_ A ^ -  + J_ B ^ - )J_p^2} \right)}},1} \right\},}\\
{{\varphi _2} = \left\{ { - \frac{{h + g}}{{2(J_ A ^ -  + J_ B ^ - ){J_p}}}, - \sqrt {\frac{{f - g}}{2}} \frac{{h - g}}{{\left( {g - hJ_B^ -  - 2(J_ A ^ -  + J_ B ^ - )J_p^2,} \right)}}} \right.,\left. { - \frac{{\sqrt 2 (J_ A ^ -  + J_{B}^ - ){J_p}\sqrt {f - g} }}{{\left( {g - hJ_ A ^ -  - 2(J_ A ^ -  + J_ B ^ - )J_p^2} \right)}},1} \right\},}\\
{\begin{array}{*{20}{l}}
{{\varphi _3} = \left\{ {\frac{{ - h + g}}{{2(J_ A ^ -  + J_ B ^ - ){J_p}}}} \right., - \sqrt {\frac{{f + g}}{2}} \frac{{h + g}}{{\left( {g - hJ_ A ^ -  + 2(J_ A ^ -  + J_ B ^ - )J_p^2} \right)}},\left. { - \frac{{\sqrt 2 (J_ A ^ -  + J_{B}^ - ){J_p}\sqrt {f + g} }}{{\left( {g + hJ_ A ^ -  + 2(J_ A ^ -  + J_ B ^ - )J_p^2} \right)}},1} \right\},}\\
{{\varphi _4} = \left\{ {\frac{{ - h + g}}{{2(J_ A ^ -  + J_ B ^ - ){J_p}}}} \right.,\sqrt {\frac{{f + g}}{2}} \frac{{h + g}}{{\left( {g + hJ_ A ^ -  + 2(J_ A ^ -  + J_ B ^ - )J_p^2} \right)}},\left. {\frac{{\sqrt 2 (J_ A ^ -  + J_{B}^ + ){J_p}\sqrt {f + g} }}{{\left( {g + hJ_ A ^ -  + 2(J_ A ^ -  + J_ B ^ - )J_p^2} \right)}},1} \right\},}
\end{array}}
\end{array}\]
where $h = J_A^ - J_A^ +  - J_B^ - J_B^ + $. Assuming $V\gg J_A^ -,J_A^ +, J_B^ -, J_B^ + $,

 one obtain
\begin{equation}
\begin{array}{l}
{E_{1,2}} \approx  - {J_p} + V,{\rm{ }}\left| {{\psi _{1,2}}} \right\rangle  \approx \left( { - \left| {{\phi _1}} \right\rangle  + \left| {{\phi _2}} \right\rangle } \right)/\sqrt 2 ;\\
{E_{3,4}} \approx {J_p} + V,{\rm{ }}\left| {{\psi _{3,4}}} \right\rangle  \approx \left( {  \left| {{\phi _1}} \right\rangle  + \left| {{\phi _2}} \right\rangle } \right)/\sqrt 2 .
\end{array}
\end{equation}
From these results, we observe that the density of eigenstates of the emerged clusters are localized at the left side, i.e. two particles on the left side and one particle on the right side. This picture is consistent with our numerical results for $L>2$. 

\section{Distribution characteristics for $N=2$ and $N=3$}
\begin{figure}[htbp]
\centering
\includegraphics[height=9.1cm]{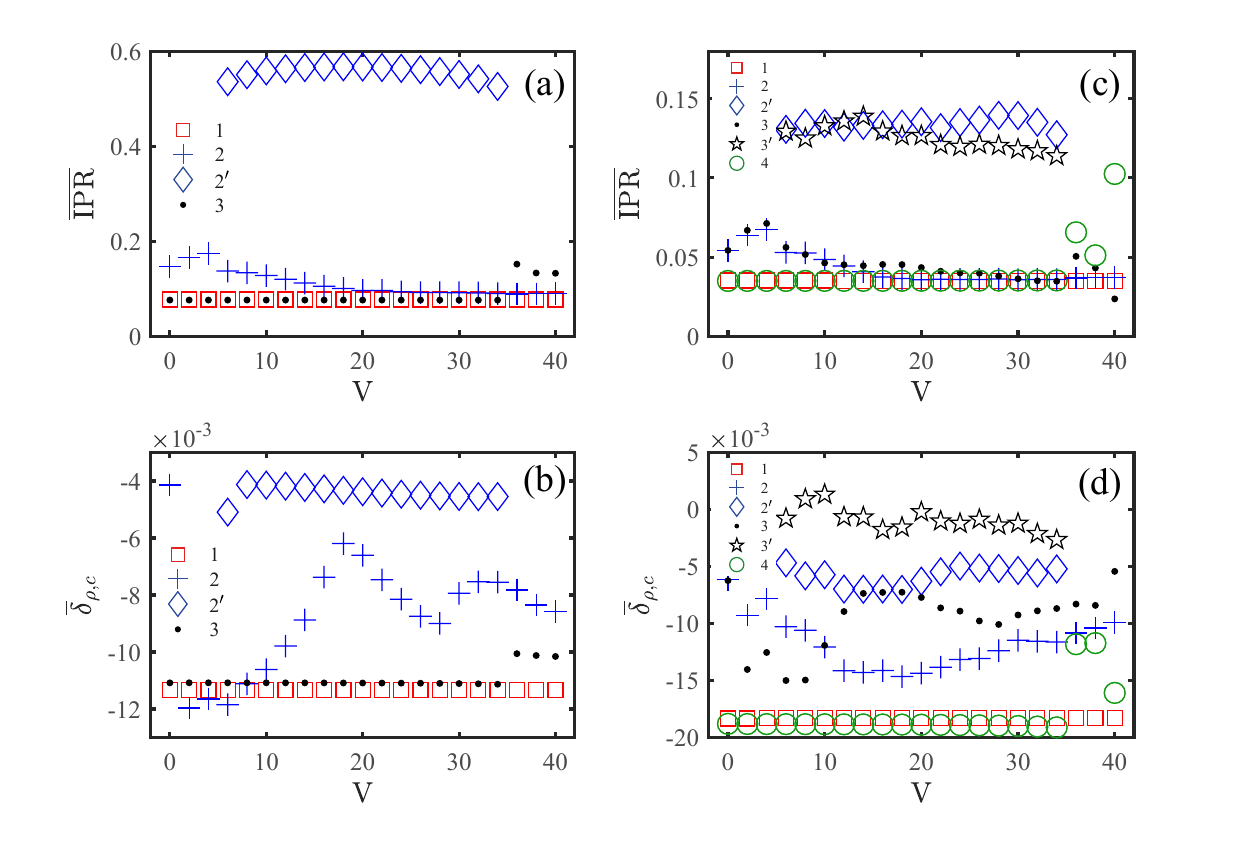}
\caption{\label{fig_IPR23}
(a) Mean IPR of each cluster when changing $V$ for $N=2$. (b) Mean density difference between the two sublattices of each cluster when changing $V$.
(c) and (d) the same as (a) and (b), but with $N=3$.
Red squares, blue crosses, black dots, and green circles correspond to the four primary clusters, and blue diamonds and black pentagrams correspond to the two secondary clusters, respectively. The system's size is chosen to be $L=10$.
Other parameters are the same with Fig.~1 in the main text.
}
\end{figure}
In this section we provide some numerical results of the distinguished behaviors of primary and secondary clusters.
As discussed in the main text, 
states with paired particles from the secondary clusters manifest the right-NHSE, which is expected to be more localized as they correspond to the non-reciprocity of each sublattice, without suffering from the destructive interference.
This is verified through the inverse participation ratios (IPR),
defined as  
\begin{equation}
{\rm IPR}_{m}=\sum_{|{n}_s\rangle}|\langle{n}_s|\psi_{m}\rangle|^4,
\end{equation}
where $\{|{n}_s\rangle\}$ is the basis of $N$-particle Fock space.
The mean IPR in each cluster is $ \overline{\rm IPR}_c  =\frac{1}{M_c} \sum_{m \in c }{\rm IPR}_m$, with $m \in c$ the index for energy in $c$ cluster and $M_c$ the Hilbert space dimension of $c$ cluster. 
We observe that this emerged secondary cluster has bigger $\overline{\rm IPR}$ than other clusters, as shown in Fig.~\ref{fig_IPR23}(a) and (c).
In Fig.~\ref{fig_IPR23}(b) and (d), we demonstrate the average density difference of the two sublattices, defined as  
\begin{equation}
\bar{\delta}_{\rho,c}=\frac{1}{M_{c}} \sum_{m\in c}\sum_{l=1}^L\left(\left|\psi_{l, m}^A\right|^2-\left|\psi_{l, m}^B\right|^2\right),
\end{equation} 
which generally takes smaller values for secondary clusters, as states with paired particles occupying the same unit cell shall have balanced densities on the two sublattices.

\section{Numerical results for $N=4$ and $N=5$}
Analytic solution of many particles problem becomes far more sophisticated for large $N$. In this section we show some numerical results about $N=4$ and $N=5$ to verify our statements in the main text. We use $V=15$ to avoid the emerged clusters overlapping with other primary clusters. When $N=4$, five primary clusters indexed by I, II, IX, VII, IX with red color and four emerged secondary clusters indexed by III, V, VI, VIII with green and blue color are observed in Fig.~\ref{fig_N4E},
numbered according to their real eigenenergies.
For $N=4$, we find that the first secondary cluster emerges from the second primary cluster, the second and third secondary clusters emerge from the third primary cluster, and the fourth secondary cluster emerges from the third primary cluster respectively, as shown in Fig.~\ref{fig_N4E}. The energy center of each cluster is given in Table.~\ref{TabE}. 
The third secondary cluster (cluster VI) with energy $E\approx 2V$ shows a left-NHSE, which indicates that the four particles form two pairs and occupy two sites,
excluding the intracell hopping $J_p$. 
On the contrary, other three secondary clusters with energy $-2J_p+V$, $V$ and $2J_p+V$ show a bipolar-NHSE. 
The mechanism is the same as the bipolar-NHSE for $N=3:$
two paired particles occupying the same unit cell forbids intracell hopping and accumulate to the left side, while the other two particles occupy different unit cells and accumulate to the right side. 
Consequently, the density distribution of these clusters is approximately the same at the two ends. 

However, things are different for $N=5$, 
where at least one particle remains unpaired. Thus the emerged secondary clusters always show a bipolar-NHSE with asymmetric density distribution, as can be seen from Fig.~\ref{fig_N5E}. The energy of each cluster are listed in Table.~\ref{TabE}. 
The first, second, fourth and sixth secondary clusters indexed by III, V, VIII, XI with green colors have two paired particles and three unpaired ones, thus the density distribution shows a bipolar-NHSE with relatively small difference at the two ends. 
The third and fifth secondary clusters indexed by VI and IX with cyan colors have four paired particles (two pairs), thus show an asymmetric distribution with larger density at the left side, which is in accordance to our intuitive picture. 
In Fig.~\ref{fig_Ent45}, we demonstrate the sublattice correlation and entanglement entropy for each eigenstate with $N=4$ and $N=5$.
The occupation configurations with different numbers of paired particles can be clearly seen from the sublattice correlation $C_n=0$, $1$, and $2$ for different clusters.
The entanglement entropy also become smaller for secondary clusters, especially those with $C_n=2$.

\begin{figure}[htbp]
\centering
\includegraphics[width=0.7\linewidth]{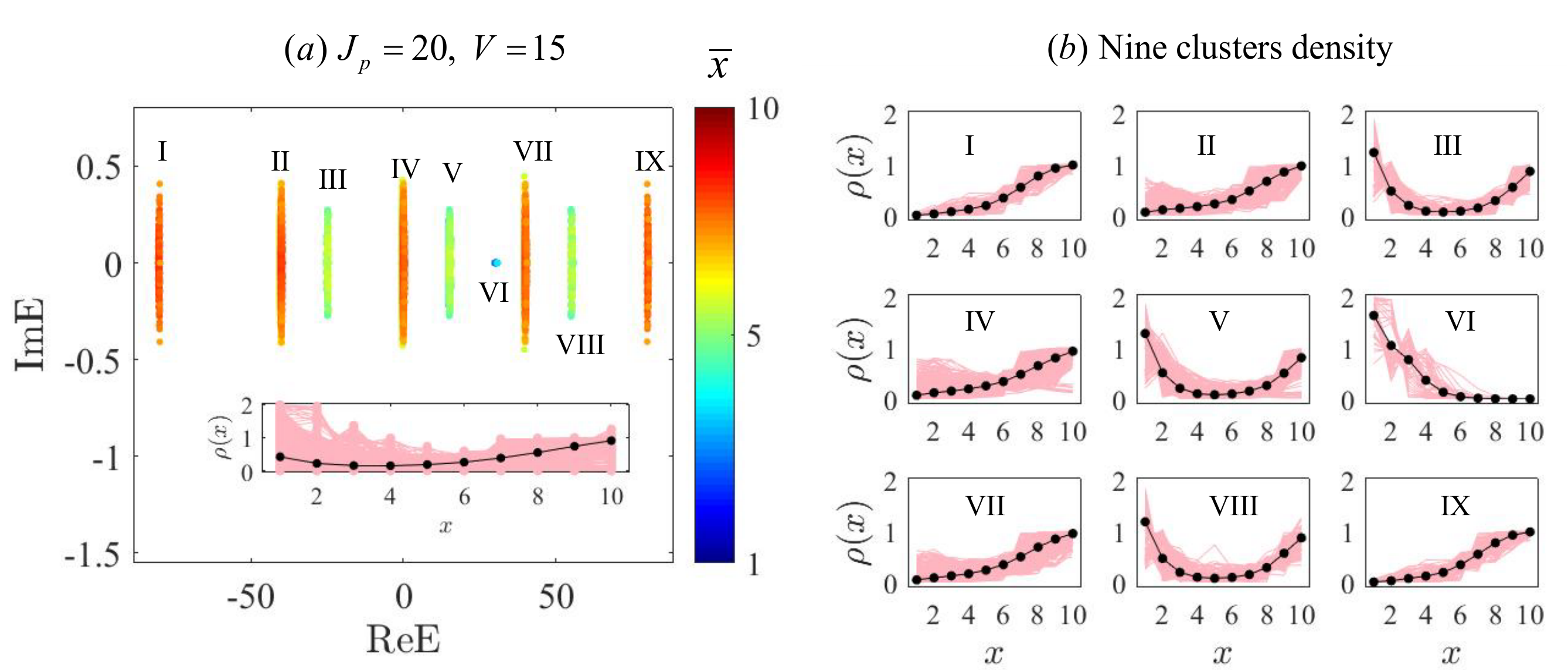}
\caption{\label{fig_N4E} 
(a) OBC energy spectrum for $N=4$, with the mean position of each eigenstate $\bar{x}$ marked by different colors.
Inset displays the density profile of each eigenstate (pink) and its average of all eigenstates (black).
(b) Density profiles and their average of each cluster. 
Right-NHSE is observed for the five primary clusters (I, II, IV, VII, and IX).
Secondary clusters show different localization properties, 
namely bipolar-NHSE for clusters III, V, and VIII, and left-NHSE for cluster VI.
Other parameters are the same as in Fig. 2 in the main text, except for $V=15$.}
\end{figure}

\begin{figure}[htbp]
\centering
\includegraphics[width=0.7\linewidth]{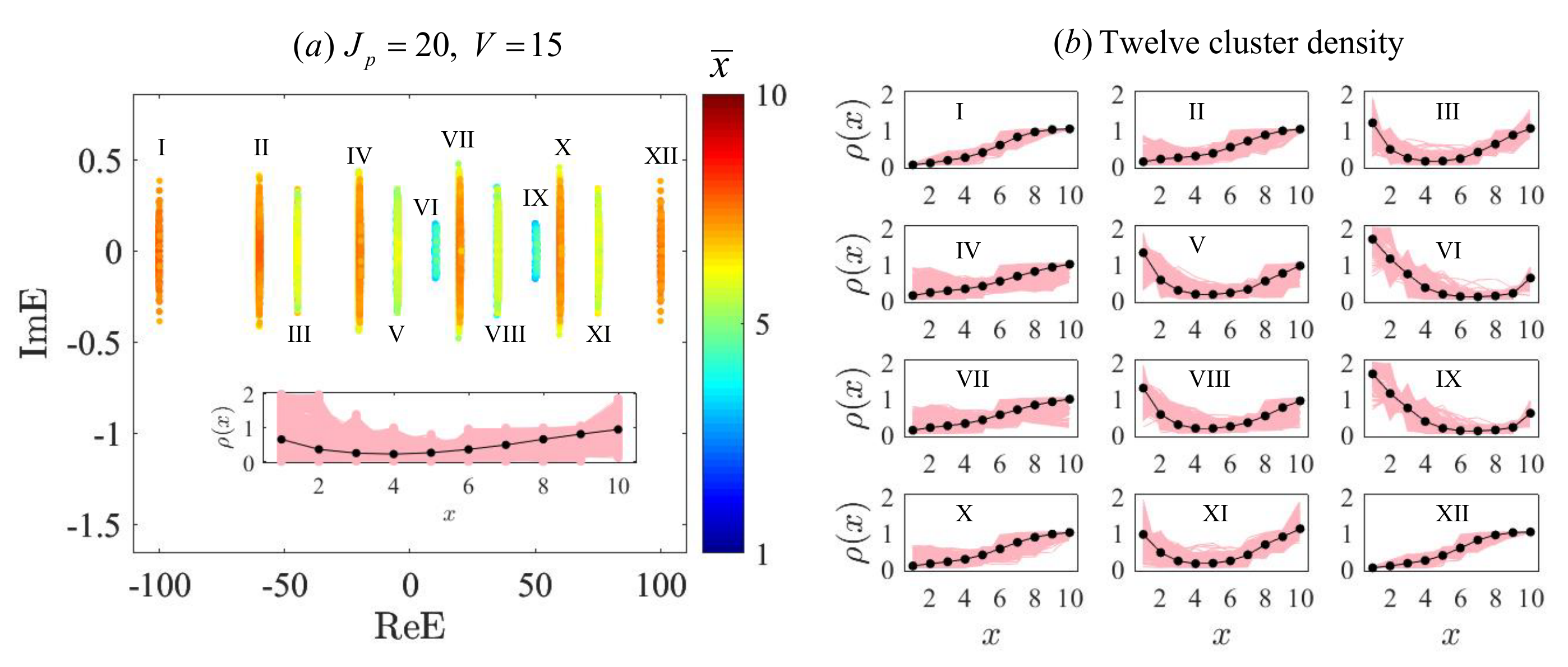}
\caption{\label{fig_N5E}
(a) OBC energy spectrum for $N=5$, with the mean position of each eigenstate $\bar{x}$ marked by different colors.
Inset displays the density profile of each eigenstate (pink) and its average of all eigenstates (black).
(b) Density profiles and their average of each cluster. 
Right-NHSE is observed for the six primary clusters (I, II, IV, VII, X, and XII).
All secondary clusters exhibit bipolar-NHSE, yet localization on the right end is weaker for clusters VI and IX.
Other parameters are the same as in Fig. 2 in the main text, except for $V=15$.
}
\end{figure}

\begin{figure}[htbp]
\centering
\includegraphics[width=0.6\linewidth]{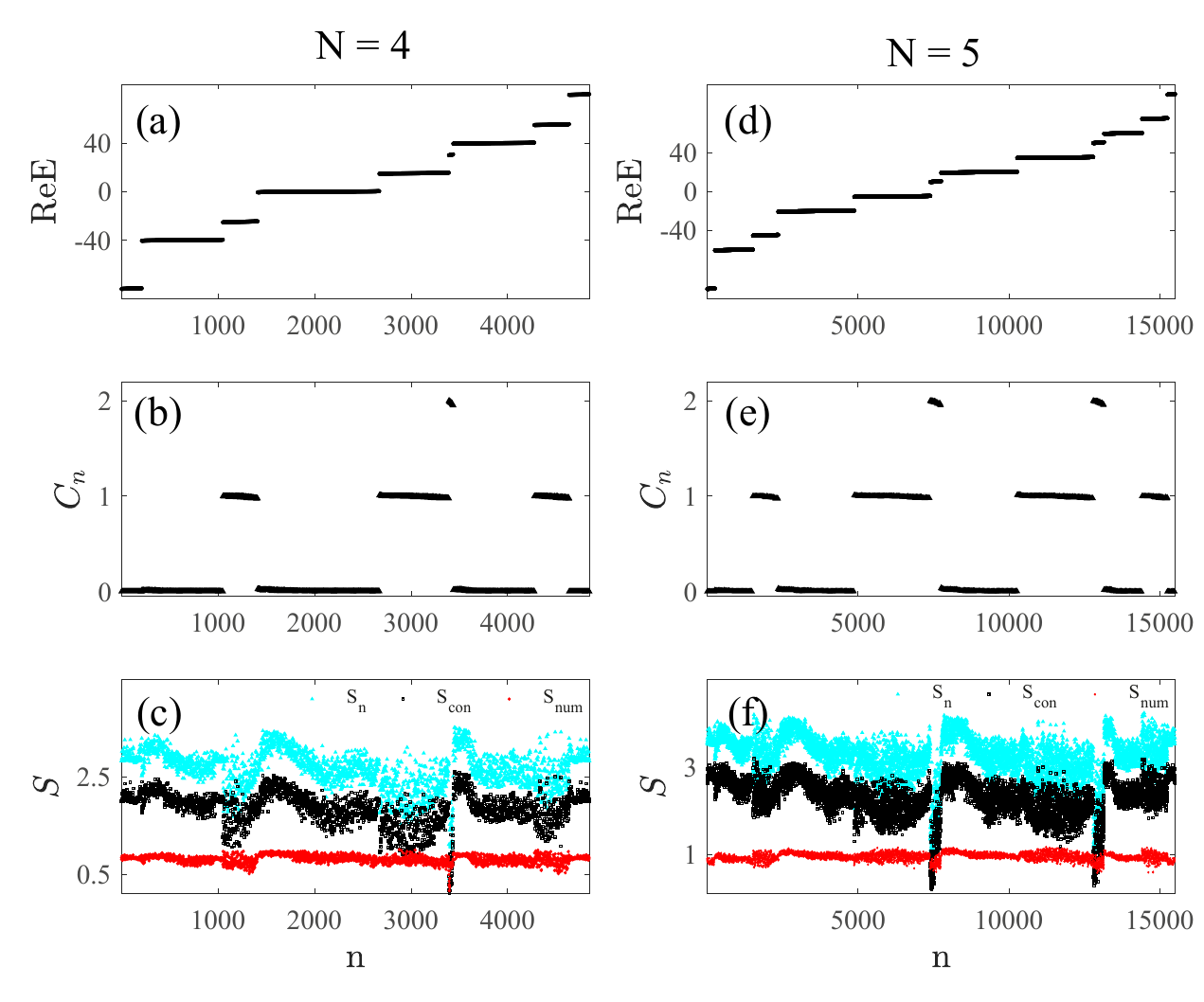}
\caption{\label{fig_Ent45}
Real eigenenergies, sublattice correlation $C_n$, and Entanglement entropy $S_n$ for $N=4$ [(a) to (c)] and $N=5$ [(d) to (f)].
The emerged secondary clusters have $C_n\approx 1$ or $C_n\approx 2$, and smaller entanglement entropy $S_n$. 
Parameters are $L=10$, $V=20$, $J_p=20$; $J_p$ and $J_{A,B}^\pm$ are the same as Fig.~1 in the main text.
}
\end{figure}


\begin{table}
\begin{center}
\begin{tabular}{|c|c|c|} \hline
$N$ &  $E$ ($V=0$) &$E$ ($V\neq0$)\\ \hline
1 & $-J_p$;  $J_p$ & $-J_p$;  $J_p$\\ \hline
2 & $-2J_p$; $0$; $2J_p$; & $-2J_p$; $0$; $V$; $2J_p$ \\ \hline
3  &$-3J_p$;  $-J_p$; $J_p$; $3J_p$; &$-3J_p$; $-J_p$; $-J_p+V$; $J_p$; $J_p+V$; $3J_p$ \\ \hline
4 & $-4J_p$; $-2J_p$; $0$; $2J_p$; $4J_p$; & $-4J_p$; $-2J_p$; $-2J_p+V$; $0$; $V$;  $2V$; $2J_p$; $2J_p+V$; $4J_p$  \\ \hline
5 & $-5J_p$;  $-3J_p$; $-J_p$; $J_p$;  $3J_p$; $5J_p$; 
& $-5J_p$; $-3J_p$;  $-3J_p+V$; $-J_p$; $-J_p+V$; $-J_p+2V$; $J_p$; $J_p+V$; $J_p+2V$; $3J_p$; $3J_p+V$; $5J_p$
\\ \hline
\end{tabular}
\caption{The approximate real energy of different clusters when $J_p\gg J_{A,B}^{\pm}$.}
\label{TabE} 
\end{center}
\end{table}

\section{Long time dynamics}
Due to the non-Hermitian gain and loss, long-time evolution behaves differently compared to the short-time evolution discussed in the main text. Here we verify the statement that the long time evolution is dominated by the state(s) with largest imaginary eigenenergy, which belongs to primary clusters that show right-NHSE. The numerical results are shown in Fig.~\ref{fig_Longtime}, 
where all initial states (paired or unpaired, $N=2$ or $N=3$) eventually evolves to the right.

\begin{figure}[htbp]
\centering
\includegraphics[width=0.4\linewidth]{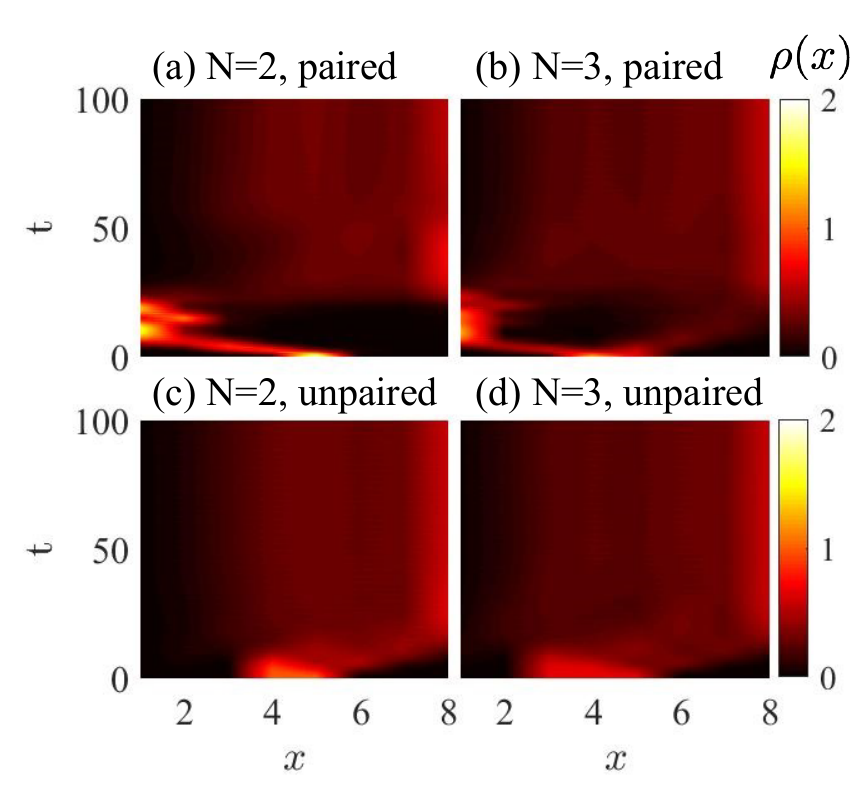}
\caption{\label{fig_Longtime}  
Long time evolution for paired and unpaired initial states.
(a) to (d) illustrate the density profiles of the evolved state $|\psi(t)\rangle$, for different initial states $|\psi(0)\rangle$
given by
(a) $|(AB)_5\rangle$, 
(b) $|(AB)_4A_5\rangle$, 
(c) $|B_4B_5\rangle$, and 
(d) $|B_3B_4B_5\rangle$,
with $J_p=20$, $V=5$, and $L=8$.
Paired and unpaired particles are seen to move to opposite directions. Long-time evolution of the states are dominated by right-NHSE.
Other parameters no mentioned here are the same as in Fig.~1 in the main text.}
\end{figure}

Other than the wavepacket dynamics, the entanglement entropy of our system shall also exhibit different behaviors in long-time and short time evolutions, as it measures how quasiparticles spread and entangle a quantum state~\cite{DeChiara2006S,Calabrese2005S,Vincenzo2017S,Orito2022S}. 
To see this, we divide the our model into two half, with unit cells $1$ to $L/2$ being the left half $\mathcal{A}$, and unit cells $L/2+1$ to $L$ being the right half $\mathcal{B}$. 
After tracing out the degree of freedoms of part $\mathcal{B}$, we obtain the entanglement entropy $S^\mathcal{A}(t)$ the subsystem $\mathcal{A}$. 

\begin{figure*}[htbp]
\centering
\includegraphics[height=7.6cm]{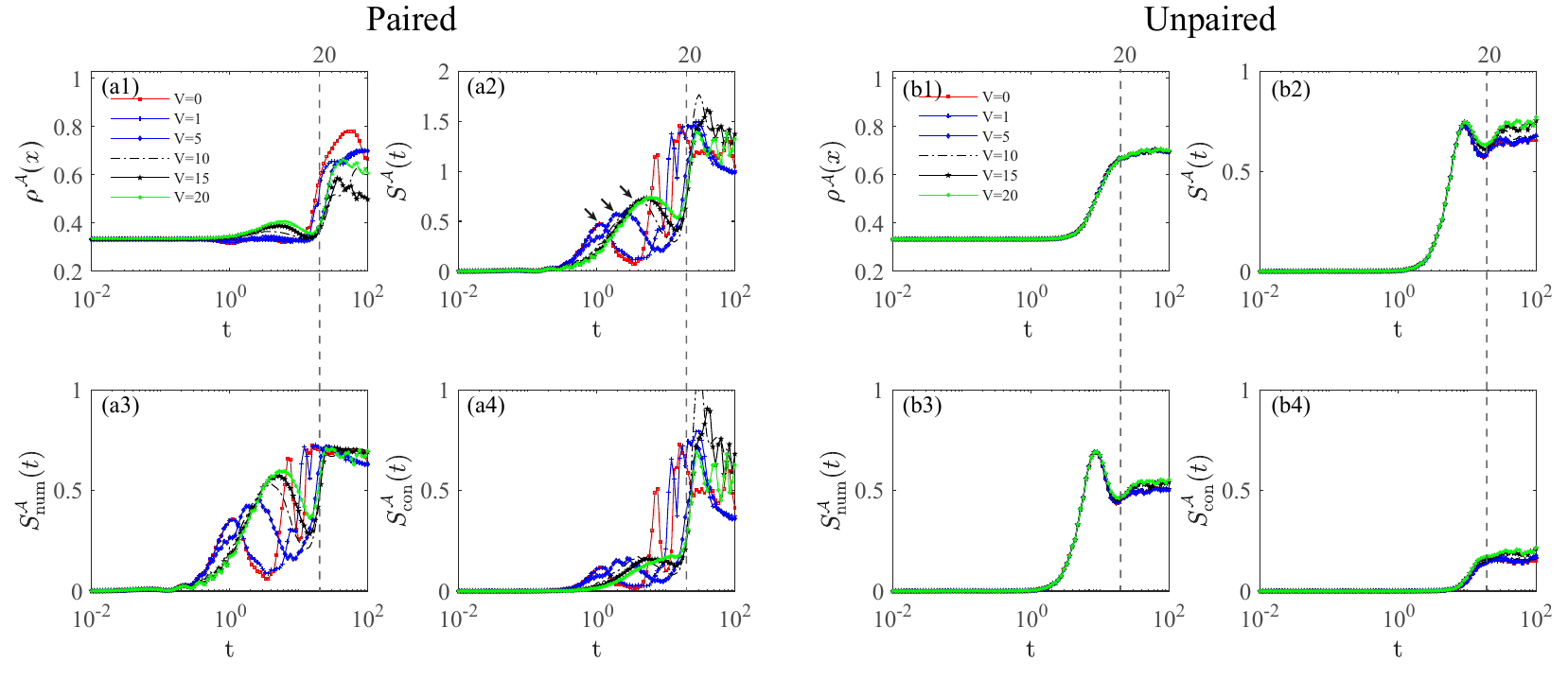}
\caption{\label{fig_numcon} Subsystem density and entanglement entropy evolution with different interaction strengths $V$, for doubly and singly occupied initial states with $N=3$. (a1) to (a4) display the subsystem density $\rho^\mathcal{A}(x)$, total entanglement entropy $S^\mathcal{A}(t)$, number entropy $S^\mathcal{A}_{\rm num}(t)$, and configuration entropy $S^\mathcal{A}_{\rm con}(t)$ respectively, for doubly occupied initial state $| (AB)_4A_5\rangle$.
(b1) to (b4) display the same quantities for singly occupied initial state  $| B_3B_4B_5\rangle$.  
Different marks represent different interaction strengths, as denoted in the figure.
$L=8$, $J_p=20$ are chosen for both cases, with other parameters being the same as in Fig. 1 in the main text. }
\end{figure*}  

Our numerical results of the entanglement entropy for doubly and singly occupied initial states are shown in Fig.~\ref{fig_numcon}, where panels (a1) and (b1) display the total density in the subsystem $\mathcal{A}$ of the evolved state, $\rho^\mathcal{A}(t)=\sum_{x\in \mathcal{A}}\langle \psi(t)|\hat{n}(x)|\psi(t)\rangle$.
As seen in Fig.~\ref{fig_numcon} (a2), entanglement entropy for a doubly occupied initial state first increases to a local maximum (pointed by the arrows), and oscillates rapidly afterward. The oscillating region roughly coincides with the region where $\rho^\mathcal{A}(t)$ becomes non-vanishing in Fig.~\ref{fig_numcon} (a1).
To understand this, note that even a doubly occupied initial state roughly keep the two particles paired in the same unit cell when moving to the left during a short-time evolution, 
it inevitably diffuses and becomes a superposition of different Fock states, thus acquire a nonzero $S^\mathcal{A}(t)$. The local maximum appears when the state hits the left boundary, after which the state tends to be more localized in the Fock space (a Fock state with two particles localized at the left end).
However, During a long-time evolution, right-NHSE of unpaired particles become the dominating factor as they correspond to larger imaginary eigeneneriges (see Fig. 2 in the main text). Consequently,  it greatly changes the dynamics of the initially paired particles,
leading to a dynamically unstable state and the rapid oscillation of entanglement entropy.
As can be seen in Fig. \ref{fig_numcon} (a3) and  (a4), for $t\gtrsim20$, the number entropy $S^\mathcal{A}_{\rm num}(t)$ remains roughly unchanged with the subsystem density $\rho^\mathcal{A}(t)$, and the oscillation of $S^\mathcal{A}(t)$ is mainly contributed by the configuration entropy $S^\mathcal{A}_{\rm con}(t)$.
In other words, for a long-time evolution with $t\sim 100$, a paired initial state eventually evolves to the right end of the system, 
yet its configuration of particle occupation becomes rather unstable, due to the competition between different types of NHSE for different paired and unpaired particles. 

In Fig.~\ref{fig_numcon} (b), we display the same quantities as in panel (a), but with singly occupied initial state $| B_3B_4B_5\rangle$. An interaction-independent local maximum of entanglement entropy is seen when $t<20$, 
as now the particles (governed by right-NHSE) are unpaired and immune to the interaction.
Similar to the previous case, 
the dominating factor of time evolution changes from maximally overlapped Fock states to eigenstates with the largest imaginary eigenenergies,
thus the long-time dynamics (with $t\gtrsim20$) of the singly occupied initial state also shows an oscillation.
However, it is much weaker compared to Fig.~\ref{fig_numcon} (a), since now 
both the initial state and 
the eigenstates with the largest imaginary eigenenergies live in the same sub-Hilbert space of
 singly occupied 
states with unpaired particles.

\section{Results for opposite non-reciprocal accumulation of different sublattice species}
In the main text, we study hardcore bosons loaded in a 1D non-Hermitian ladder lattice with the same non-reciprocal pumping direction for different sublattice species,
and unveil the separation of paired and unpaired particles in real space. 
In this section, we consider another parameter regime where the two sublattice species have opposite non-reciprocal pumping directions.
In the absence of intracell hopping, the two sublattices are decoupled into two Hatano Nelson chains, with loop-like spectrum under periodic boundary conditions (PBCs) and skin modes localized at different ends under the open boundary conditions (OBCs), as shown in Fig.~\ref{fig_Static} (a).
On the other hand, a strong intracell hopping ($J_p=20$ in our example) couples the two sublattices, and NHSE disappears as their opposite non-reciprocal pumps cancel each other, as shown in Fig.~\ref{fig_Static} (b).
\begin{figure}[htbp]
\centering
\includegraphics[width=0.55\linewidth]{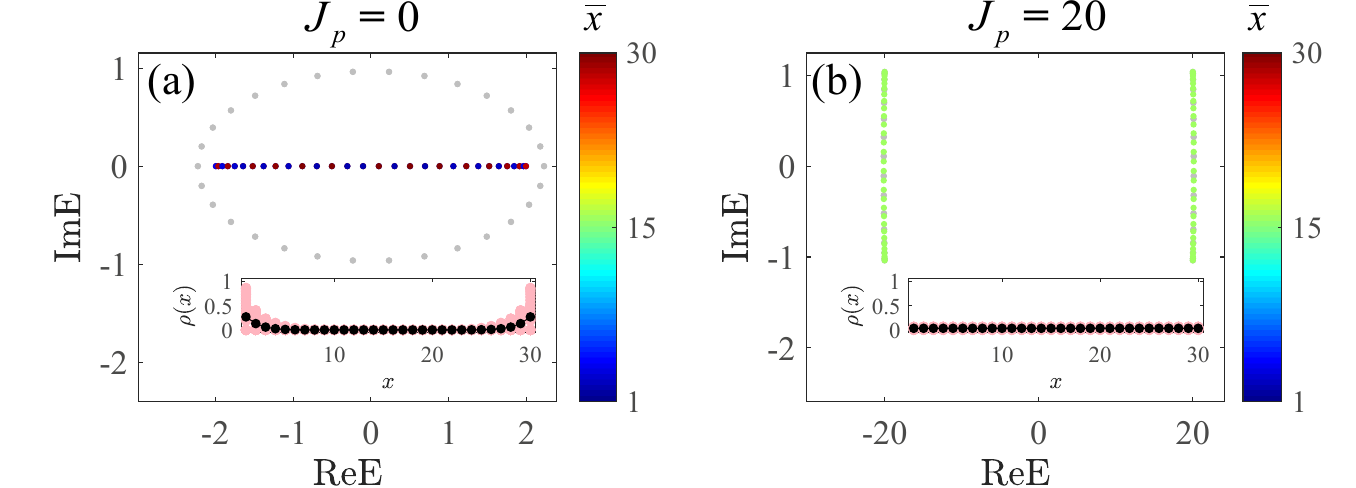}
\caption{\label{fig_Static}
Static properties for the 1D non-Hermitian ladder with opposite localization directions.  Gray (colored) dots indicate the PBC (OBC)  spectrum,  
with different colors representing the mean position of each state $\overline{x}=\langle \hat{x}\rangle$.
Insets shows the OBC densitiy of each eigenstate (pink) and its average of all eigenstates (black).
Intracell hopping term is $J_p=0$ in (a) and $J_p=20$ in (b). Note the red and blue dots indicating opposite localization directions overlap each other in (a). Other parameters are $J_{A}^+=0.61, J_{A}^-=1.65$, $J_{B}^+=-0.61$,$J_{B}^{-}=-1.65,L=30, N=1$.
}
\end{figure}

In the presence of multiple particles, we also expect different behaviors for paired and unpaired particles in the dynamical level, as 
intracell hopping also plays a determining role in this scenario.
In Fig.~\ref{fig_pairunpair}(a) and (c), we observe a clear sublattice-dependent unidirectional propagation with $J_p=0$,
i.e. particles move to the left in $A$ sublattice and to the right in $B$ sublattice,
for both doubly and singly occupied initial states. 
For a strong intracell hopping, on the other hand, both initial states tend to diffuse bi-directionally in the system, but still exhibit distinguished dynamical signatures.
As seen in Fig. \ref{fig_pairunpair}(b), paired particles show a weak bipolar pumping, leaving the center of the system nearly empty (with much smaller $\rho(x)$ when $t\gtrsim2$). An intuitive understanding is that the two paired particles are immune to intracell hopping in the very beginning, and experience the sublattice-dependent non-reciprocal pumping; yet their opposite pumping directions separates the two particles, hence the intracell hopping comes into play and the non-reciprocal pumping is thus suppressed.
For the other initial state with two unpaired particles, we observe that
the diffusion of the states is accompanied by a sublattice dependent oscillation
which may be caused by a different interplay between the bi-directional NHSE
at $J_p=0$
and the skin-free scenario at $J_p=20$. 
Details and mechanisms of these enigmatic dynamical phenomena remain to be further investigated.

\begin{figure}[htbp]
\centering
\includegraphics[width=0.55\linewidth]{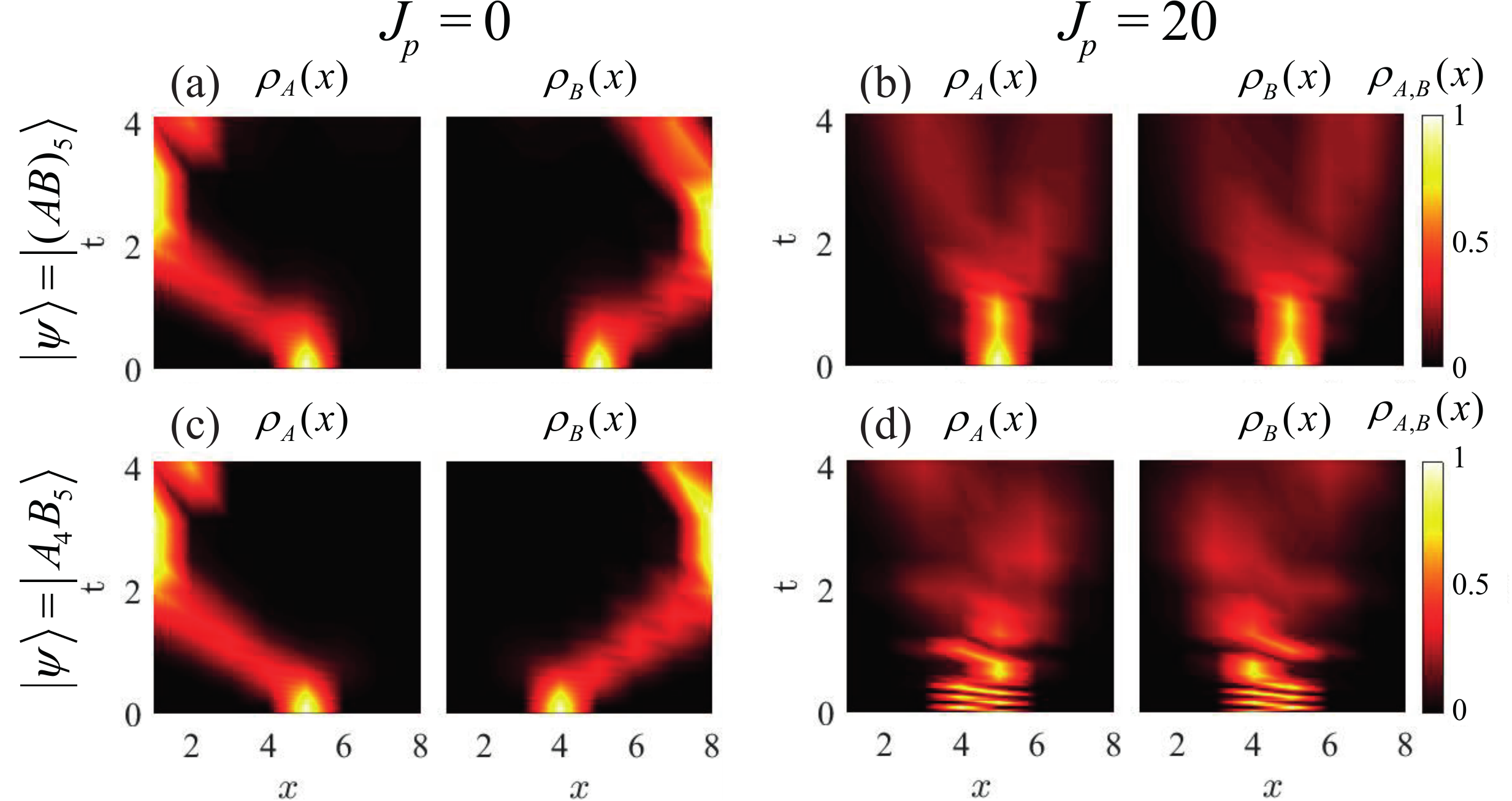}
\caption{\label{fig_pairunpair}
Dynamics for the 1D non-Hermitian ladder with sublattice-dependent non-Hermitian pumping directions. (a) and (b) Density evolutions on each sublattice of a doubly occupied initial state $| (AB)_5\rangle$ in $L=8$ lattice sites, with $J_p=0$ and $J_p=20$ respectively. (c) and (d) Density evolutions of a singly occupied initial state $| B_4A_5\rangle$ with the same parameters. 
Different colors indicate the normalized density profile.
Other parameters are $L=8$, $V=5$, $J_{A}^+=0.61$, $J_{A}^-=1.65$, $J_{B}^+=-0.61$, and $J_{B}^{-}=-1.65$.
}
\end{figure}

\section{Pairing separation in spin-1/2 fermionic chain}
In the main text, 
we have focused on the hardcore bosons loaded in a 1D non-Hermitian ladder lattice, and reveal the emergence of the occupation-dependent particle separation.
Intuitively, we expect similar phenomena to emerge in systems with different internal degree of freedom, for example,
spin-1/2 fermions with up and down spins.
However, exchange antisymmetry of fermions may also affect the static and dynamical properties.

\subsection{Interaction induced pairing-dependent many-body NHSE}
In this subsection we consider a spin-1/2 fermionic gas loaded in a 1D lattice with $L$ lattice sites with a on-site Hubbard interaction, whose Hamiltonian takes the same form as Eq. 1 in the main text, but with subscript indexes $A$ and $B$ replaced by $\uparrow$ and $\downarrow$ for the two spin species,
\begin{eqnarray}
\hat{H}_{\rm fermions} &=&  \sum\limits_{l=1}^{L-1}\mathbf{\hat{\psi}}^\dagger_{l+1}
\left(\begin{array}{cc}
J^+_\uparrow & 0\\
0 & J^+_\downarrow
\end{array}\right)
\mathbf{\hat{\psi}}_{l}+\mathbf{\hat{\psi}}^\dagger_{l}
\left(\begin{array}{cc}
J^-_\uparrow & 0\\
0 & J^-_\downarrow
\end{array}\right)
\mathbf{\hat{\psi}}_{l+1}
\nonumber\\
&&+\mathbf{\hat{\psi}}^\dagger_{l}
\left(\begin{array}{cc}
0 & J_p\\
J_p & 0
\end{array}\right)
\mathbf{\hat{\psi}}_{l}+ V\sum\limits_{l=1}^{L} {\hat{n}_{\uparrow,l}\hat{n}_{\downarrow,l}} .
\nonumber
\end{eqnarray}
Here $\hat{\psi}_l=(\hat{c}_{\uparrow,l},\hat{c}_{\downarrow,l})^T$ are the annihilation operators of a fermion at the site $l$ of each spin,
$J_{\uparrow,\downarrow}^\pm$ and $J_p$ the spin-dependent nearest-neighbor hopping and spin flipping amplitudes respectively,
and $V$ the strength of Hubbard interaction. 

The eigensolutions of this model for $N=2$ and $N=3$ fermions are demonstrated in Fig. \ref{fig_N2N3}.
As shown in Fig. \ref{fig_N2N3}(a) and (b) for $N=2$ particles, when the interaction is turned off, 
all many-body eigenstates accumulate to the left and possess real eigenenergies when spin flipping is switched off ($J_p=0$), 
and to the right in the presence of a strong spin flipping ($J_p=20$) that induces the direction reversal of NHSE.
Notably, the deactivation of direction reversal does not occur in the absence of interaction, and each eigenstate in Fig. \ref{fig_N2N3}(b) exhibit the right-NHSE as in the single-particle level,
in sharp contrast to the hardcore boson case for the same parameters [see Fig. 2(b) in the main text]. 
However, a strong interaction still induces a secondary cluster with left-NHSE, as shown in Fig. \ref{fig_N2N3}(c) and (d) for $V=20$.
The coexistence of left- and right-NHSE can be attributed to different pairing conditions of many-body eigenstates, giving raise to a spatial separation of paired and unpaired particles.
namely, paired fermions with opposite spin forbid spin flipping due to the quantum degenerate pressure, in analogous to the paired hardcore bosons discussed that forbid intracell hopping, as discussed in the main text. 
Similarly, the four primary eigenstate clusters for $N=3$ with $J_p=20$ and $V=0$ possess only right-NHSE, and bipolar-NHSE only emerges in the two secondary clusters induced by a strong interaction, as showin in  Fig. \ref{fig_N2N3}(e) to (h).
These results suggest that interaction plays a much more important role to induce occpuation/pairing-dependent particle separation for fermions, compared with their bosonic analogs.

\begin{figure*}
\includegraphics[width=1\linewidth]{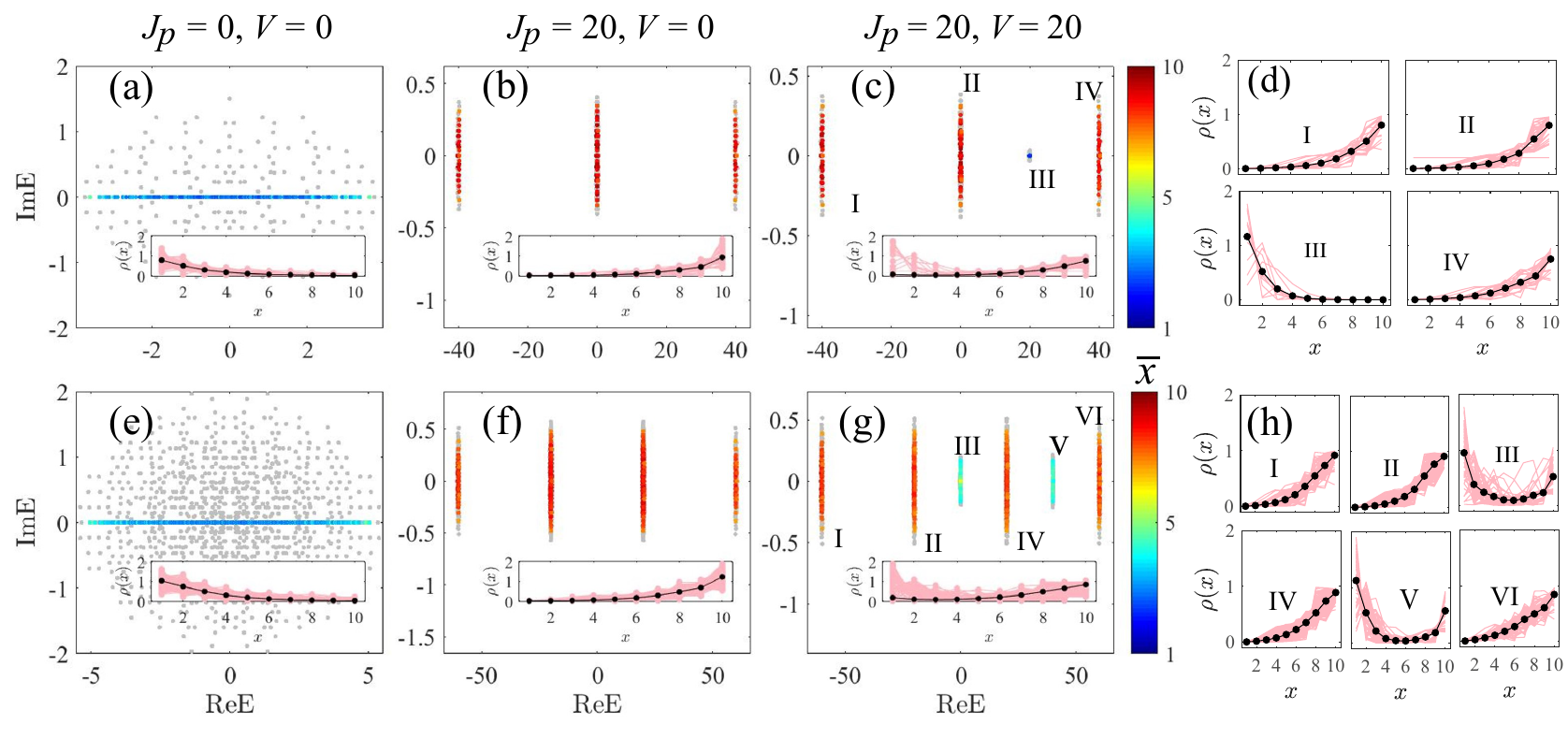}
\caption{Energy spectrum and skin accumulations for $N=2$ and $N=3$. 
For both cases, gray and colored dots indicate energy spectra under PBCs and OBCs respectively, with the colormap representing the mean position of each eigenstate, $\overline{x}=\langle \hat{x}\rangle$.
(a) to (c) Energy spectra for $N=2$ for different values of $J_p$ and $V$, with insets showing the density profiles of all eigenstates (pink) and their average (black).
(d) Density distributions of the four clusters in (c).
Three primary eigenstate clusters with right-NHSE are seen to be induced by the spin flipping $J_p$, and a secondary one with left-NHSE is separated from them when turning on the interaction $V$.
(e) to (h) the same quantities as in (a) to (d) for our system with $N=3$ particles, which hosts four primary clusters with right-NHSE, and
two secondary clusters exhibiting bipolar-type of skin localization.
$L=10$ is chosen for all panels.
Other parameters are $J_{A}^+=0.45, J_{A}^-=1.24$, $J_{B}^+=-0.82$,$J_{B}^{-}=-1.22$, which are the same as in Fig. 1 in the main text.}
\label{fig_N2N3}
\end{figure*}

\subsection{Dynamic properties}
We next examine the dynamical properties for the spin-1/2 fermions in our system.
In contrast with hardcore bosons, pairing-dependent NHSE for fermions requires a strong interaction, and so should particle separation.
As shown in Fig. \ref{fig_D}(a), both paired and unpaired particles move to the right during the time evolution, reflecting the right-NHSE induced by a strong flipping. When gradually increasing the interaction strength, the right-NHSE is suppressed for a spin pair, which is frozen at its initial position for a short period of time, and diffuses afterward, as shown in Fig. \ref{fig_D}(b) to (d).
Finally, when the interaction is strong enough, paired particles are pumped to the left side of the lattice, and are thus separated from unpaired one, as shown in Fig. \ref{fig_D}(e) and (f) for $V=15$ and $V=20$ respectively.
On the hand, the right-NHSE of unpaired particles is also suppressed by the interaction, which shows a stronger diffusion for stronger interaction strength $V$.
To conclude, the pairing-dependent particle separation of fermions is also more sensitive to interaction in the dynamical level,
which awaits further explorations.

\begin{figure}
\centering
\includegraphics[width=0.8\linewidth]{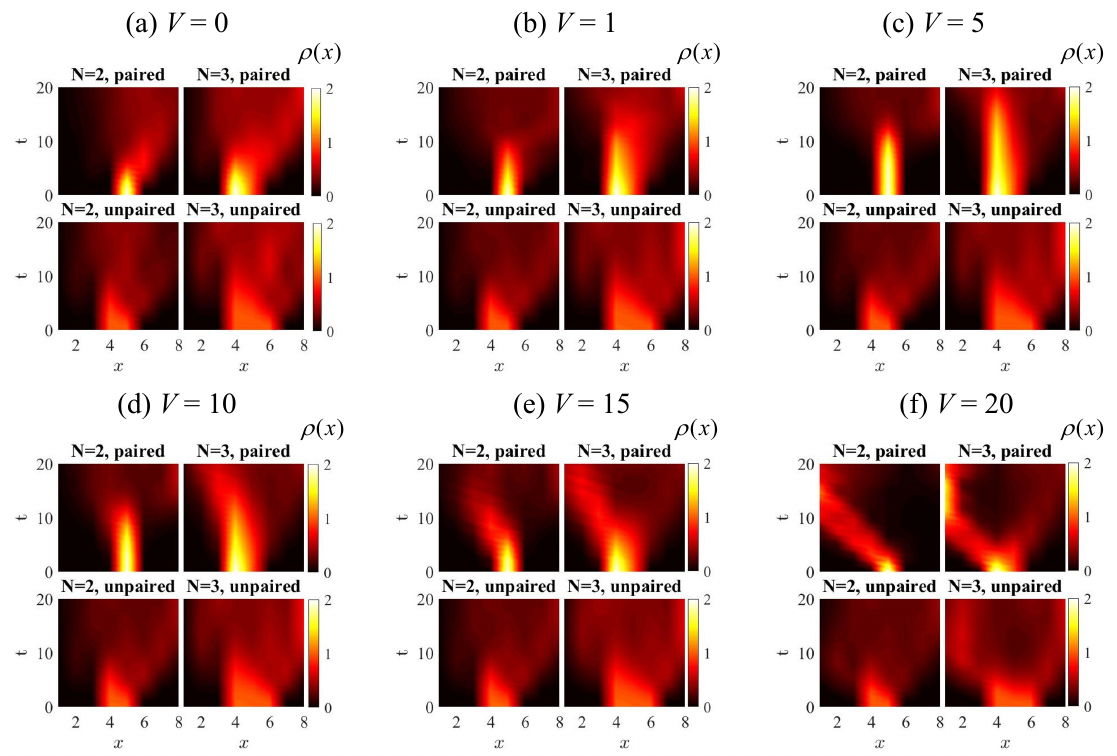}
\caption{\label{fig_D}
Dynamics evolution for paired and unpaired initial states of spin-1/2 fermions.
(a) to (f) illustrate the density profiles of the evolved state $|\psi(t)\rangle$, for different initial states $|\psi(0)\rangle$
given by
$N=2$ paired $|(\uparrow\downarrow)_5\rangle$, 
$N=3$ paired $|(\uparrow\downarrow)_4\uparrow_5\rangle$, 
$N=2$ unpaired $|\downarrow_4\downarrow_5\rangle$, and 
$N=3$ unpaired $|\downarrow_4\downarrow_5\downarrow_6\rangle$,
with $J_p=12$, (a)$V=0$,(b)$V=1$,c$V=5$,(d)$V=10$, (e)$V=15$, (f)$V=20$ and  $L=8$.
Paired and unpaired particles are seen to move to opposite directions for sufficient large $V$.
The parameters no mentioned here are $J_{A}^+=0.45, J_{A}^-=1.24$, $J_{B}^+=-0.82$,$J_{B}^{-}=-1.22$, which are the same as in Fig. 1 in the main text.}
\end{figure}

\end{widetext}

\end{document}